\documentclass[twocolumn,nofootinbib]{revtex4}

\usepackage{epsfig}
\usepackage{amsmath}
\usepackage{amsfonts}
\usepackage{amssymb}
\usepackage{graphicx}
\usepackage[hyperindex=true,colorlinks=true,pagebackref=false,citecolor=blue,plainpages=false,pdfpagelabels,breaklinks=true,linkcolor=blue,urlcolor=blue,
runcolor=blue,anchorcolor=blue]{hyperref}

\usepackage[T1]{fontenc}
\usepackage[utf8]{inputenc}

\begin{document}
%%%%%%%%%%%%%%%%%%%%%%%%%%%%%%%
\title{{\bf Geodesic motion in a stationary dihole spacetime}}
%%%%%%%%%%%%%%%%%%%%%%%%%%%%%%%
\author{F. L. Dubeibe}
\email{fldubeibem@unal.edu.co}
\affiliation{Facultad de Ciencias Humanas y de la Educaci\'on, Universidad de los Llanos, Villavicencio, Colombia}
\affiliation{Grupo de Investigaci\'on en Relatividad y Gravitaci\'on, Escuela de F\'isica, Universidad Industrial de Santander, A.A. 678, Bucaramanga, Colombia  \\}

\author{Jos\'e D. Sanabria-G\'omez}
\email{jsanabri@uis.edu.co} \affiliation{Grupo de Investigaci\'on en Relatividad y Gravitaci\'on, Escuela de F\'isica, Universidad Industrial de Santander, A.A. 678, Bucaramanga, Colombia}

\date{\today}
%%%%%%%%%%%%%%%%%%%%%%%%%%%%%%
\begin{abstract}
The knowledge of the properties of the different exact solutions modeling binary systems, is a necessary step towards the classification of physically suitable solutions and its corresponding limits of applicability. In the present paper, we perform an analysis of the geodesics around two counter--rotating Kerr--Newman black holes endowed with opposite electric charges, which achieve equilibrium by means of a strut between their constituents. We find that bounded and unbounded orbits are possible. However, test particles may cross between the black holes only if their angular momentum equals zero, otherwise, there exist a repulsive potential, which prohibits such orbits. Two important aspects are pointed out for these trajectories: ({\it i}) the motion of photons is affected once crossing the strut; and ({\it ii}) massive particles exhibit oscillatory motion, as a first analog of the Sitnikov problem in general relativity. The radius of the innermost stable circular orbit as a function of the physical parameters of the black holes is also investigated.
\end{abstract}


\maketitle
%%%%%%%%%%%%%%%%%%%%%%%%%%%%%
\section{Introduction}\label{Introduction}
%%%%%%%%%%%%%%%%%%%%%%%%%%%%
Interacting black holes were one of the first proposed possible candidates for detection of gravitational waves. But even after its detection by the LIGO collaboration \cite{LIGO}, a suitable astrophysical exact solution of Einstein and  Einstein-Maxwell equations to model a binary system of black holes, still not found. However, since the 40's some attempts have been made (see \cite{MAJUMDAR, HARTLE, KRAMER, BONNOR}). In order to ensure the system remains in equilibrium, some mechanism have been developed. For example, the Majumdar--Papapetrou solutions are in equilibrium as a consequence of the balance between their electrostatic repulsion and gravitational attraction, while in the multi-black holes case (see {\it e.g} \cite{ISRAEL_KAN}), the equilibrium state can only be reached by means of conical singularities along the symmetry axis, so-called Weyl struts.

In 2000 Emparan \cite{EMPARAN} presented and analyzed an exact solution of the Einstein--Maxwell equations for static pairs of black holes. In this case the suspended equilibrium is due to magnetic external fields, removing the strut. He coined the term "dihole'' to mean two separated objects carrying opposite electric or magnetic charges and therefore possessing a magnetic dipole moment. In a later paper, Emparan and Teo \cite{EMPARAN_TEO} considered a stationary axisymmetric solution presented by Manko {\it et. al} \cite{MANKO_MARTIN_RUIZ}, interpreting it as a non--extremal dihole solution. They also generalized the Manko's results to a solution of Einstein--Maxwell--dilaton theory and to the $U(1)^4$ theories coming from the compactified string/M-theory. Some years later, Cazares {\it et. al} \cite{CAZARES-COMPEAN_MANKO} reparametrized the Emparan-Teo non-extremal black dihole solution in terms of Komar quantities and the distance between the objects, in order to obtain a simpler form for the metric and a more intuitive physical representation of the two body configuration. 

Recently, Manko {\it et. al} \cite{Manko} found the simplest model for stationary black diholes (hereafter SSBD) consisting of two counter-rotating Kerr--Newman black holes endowed with opposite electric charges. The solution was built setting the magnetic monopole to zero, such that the solution possesses electric and magnetic dipoles and constitutes a generalization to the diholes proposed by Emparan \cite{EMPARAN}. It should be pointed out that the SSBD solution does not have ring singularities but, nevertheless presents a strut separating the two constituents. Moreover, the SSBD solution satisfies the Gabach--Clement inequality for interacting black holes with struts \cite{GABACH}. 

In a different context, the necessity of suitable initial data for the numerical simulations of binary systems with nearly extremal spins, requires, for example, the knowledge of quasiequilibrium data based on the superposition of Kerr--Schild metrics \cite{LOVELACE}. Therefore, the information provided by exact solutions could be useful for the characterization of more realistic binary systems. With this in mind, in the present paper we explore the external gravitational field of a binary configuration of counterrotating Kerr--Newman black holes endowed with opposite electric charges ($\sigma \in \mathbb{R}$), through the study of geodesics in the SSBD solution \cite{Manko}.

The paper is organized as follows: In section \ref{sec2} we explicitly present the SSBD solution and its main properties. Next, in section \ref{sec3}, we derive the equations of motion for timelike and null geodesics confined to the dihole's equatorial plane and classified the possible orbits. The dependence of the innermost stable circular orbit whit the physical parameters of the source is also investigated. Timelike and null geodesics out of the equatorial plane are considered in section \ref{sec4}. Finally, in section \ref{Conclusions} the conclusions are presented.

%%%%%%%%%%%%%%%%%%%%%%%%%%%%
\section{The Simplest Stationary Dihole Solution} \label{sec2}
%%%%%%%%%%%%%%%%%%%%%%%%%%
The SSBD solution has been obtained in the paper \cite{Manko}, and it is defined by Ernst potentials of the form

\begin{equation}
{\cal{E}}=\frac{A-B}{A+B},\quad \Phi=\frac{C}{A+B},
\end{equation}

with

\begin{widetext}
\begin{eqnarray*}
A&=&R^2(M^2 -Q^2 \nu)(R_{+}-R_{-})(r_{+}-r_{-})+4\sigma^2(M^2 +Q^2
\nu)(R_{+}-r_{+})(R_{-}-r_{-})\\&&+2R\sigma[R\sigma(R_{+}r_{-}+R_{-}r_{+})
+i M a \mu(R_{+}r_{-}-R_{-}r_{+})],\\
B&=&2 M R \sigma[R\sigma(R_{+}+R_{-}+r_{+}+r_{-})
-(2M^2-i M a\mu)(R_{+}-R_{-}-r_{+}+r_{-})],\\
C&=&2C_{0}R\sigma[(R+2\sigma)(R\sigma-2M^2-i M a\mu)
(r_{+}-R_{-})+(R-2\sigma)(R\sigma+2M^2+i M a\mu)(r_{-}-R_{+})], \nonumber\\
R_{\pm}&=&\sqrt{\rho^2+\left(z+\frac{R}{2}\pm\sigma \right)^2},
\quad r_{\pm}=\sqrt{\rho^2+\left(z-\frac{R}{2}\pm\sigma
\right)^2},
\end{eqnarray*}

where the constant quantities $\sigma$, $\mu$, $\nu$ and $C_{0}$ are defined as

\begin{eqnarray}
\sigma&=&\sqrt{M^2-\left(\frac{M^2 a^2 \left[(R+2 M)^2+4
Q^2\right]} {\left[M(R+2M)+Q^2\right]^2}+Q^2\right)\frac{R-2
M}{R+2M}}, \nonumber\\ \mu&=&\frac{R^2-4 M^2}{M(R+2M)+Q^2},\quad
\nu=\frac{R^2-4 M^2}{(R+2M)^2+4 Q^2},\quad C_{0}=-\frac{Q
\left(R^2-4 M^2+2 i M a \mu \right)}{(R+2M) \left(R^2-4
\sigma^2\right)}. \label{sigma}
\end{eqnarray}

The corresponding metric coefficients $f$, $\gamma$ and $\omega$, for the stationary axisymmetric line element

\begin{equation} \label{Papa}
ds^2 = f^{-1} [ \, e^{2 \gamma} \, (d\rho^2 + dz^2) + \rho^2 \,
d\phi^2 ] - f \, (dt - \omega d\phi)^2,
\end{equation}

are given by the expressions

\begin{equation}
f=\frac{A \bar{A}-B \bar{B}+C \bar{C}}{(A+B)(\bar{A}+\bar{B})},
\quad e^{2\gamma}=\frac{A \bar{A}-B \bar{B}+C \bar{C}}{16 R^4
\sigma^4 R_{+}R_{-}r_{+}r_{-}},\quad \omega=-\frac{{\rm
Im}[2G(\bar{A}+\bar{B})+C \bar{I}]}{A \bar{A}-B \bar{B}+C
\bar{C}},
\end{equation}

whit

\begin{eqnarray}
G&=&-z B+R\sigma\{R(2M^2-Q^2\nu)(R_{-}r_{-}-R_{+}r_{+})
+2\sigma(2M^2+Q^2\nu)(r_{+}r_{-}-R_{+}R_{-}) \nonumber\\
&&+M[(R+2\sigma)(R\sigma-2M^2+i M a
\mu)+2(R-2\sigma)Q^2\nu](R_{+}-r_{-}) \nonumber\\ &&+
M[(R-2\sigma)(R\sigma+2M^2-i M a \mu)-2(R+2\sigma)Q^2\nu](R_{-}-r_{+})\},\nonumber\\ \nonumber\\
I&=&-z C+2C_{0}M[R^2(2M^2-2\sigma^2+i M a \mu)(R_{+}r_{+}+R_{-}r_{-})
+2\sigma^2(R^2-4 M^2-2 i M a \mu)(R_{+}R_{-}+r_{+}r_{-})] \nonumber\\
&&-C_{0}(R^2-4\sigma^2)\{2M[R\sigma(R_{+}r_{-}-R_{-}r_{+})
+(2M^2+i M a\mu)(R_{+}r_{-}+R_{-}r_{+})] \nonumber\\ &&+R
\sigma[R\sigma(R_{+}+R_{-}+r_{+}+r_{-})+(6M^2+i M
a\mu)(R_{+}-R_{-}-r_{+}+r_{-})+8 M R\sigma]\}.
\end{eqnarray}
\end{widetext}

\begin{center}
\begin{figure}
\includegraphics[width=5cm,angle=0]{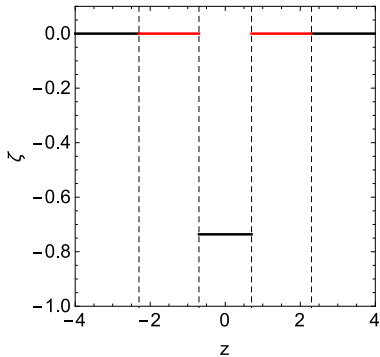}
\caption{(Color online) Limit of the metric function $\gamma(\rho, z)$ and its dependence with de $z$-coordinate. Parameters have been set to $R=3, M=a=Q=1$ (in which case $\sigma\approx 0.8$). The red lines denote the KN black holes.} \label{lgamma}
\end{figure}
\end{center}

This solution satisfies the asymptotic flatness conditions,

\begin{equation*}
\lim\limits_{\rho,z\to\infty}f(\rho,z)=1,\,\, \lim\limits_{\rho,z\rightarrow\infty}\gamma(\rho,z)=0, \,\, \lim\limits_{\rho,z\rightarrow\infty}\omega(\rho,z)=0,
\end{equation*}

while the elementary flatness condition is not satisfied in the interval $(-R/2+\sigma, R/2-\sigma)$, where 

\begin{equation*}
\lim_{\rho \to 0} \gamma(\rho,z)=\zeta(z),   
\end{equation*}

due to the existence of a strut separating the constituents. In Fig.~\ref{lgamma} we present  $\zeta=\lim_{\rho \to 0} \gamma(\rho,z)$ as a function of $z$. The red lines represents the constituents and the black lines the result for $\zeta$ when evaluated at different points of the $z$-axis.

The multipolar moments characterizing the dihole solution can be calculated using the Fodor-Hoenselaers-Perj\'es procedure \cite{HoensPerj}, taking into account the corrections made by Sotiriou and Apostolatos \cite{Sotiriou2004}, yielding

\begin{eqnarray}
M_0 &=& 2 M, \, M_1=0, \, M_2=M(R^2-8 M^2+4\sigma^2)/2,\nonumber\\
J_0 &=& 0,\, J_1=0,\, J_2=2 M^2 a \mu,\, Q_0=0,\nonumber\\
Q_1&=&Q(2M-R),\, Q_2=0,
B_0=0,\nonumber\\
B_1&=&2 a M Q\mu/(R+2M),\, B_2=0,\nonumber
\end{eqnarray}

whence it follows that $M$ and $Q$ denote the mass and the electric charge of each constituent, $R$ represents the coordinate separation between the sources and $a=J/M$ is the angular momentum per unit mass.

\begin{figure}
\centering
\includegraphics[width=3.5cm,angle=0]{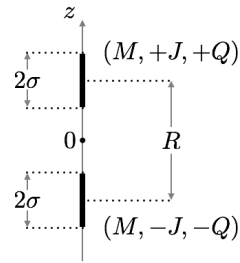}
\caption{Schematic representation of the black dihole configuration of two Kerr--Newman black holes ($\sigma \in \mathbb{R}$). The solid bold lines denote the location of the sources on the symmetry axis.} \label{fig1}
\end{figure}

Depending on the parameter $\sigma$, the configuration describes either two Kerr--Newman black holes (real $\sigma$) or two superextreme Kerr--Newman constituents (pure imaginary $\sigma$). In Fig.~\ref{fig1}, we show an schematic representation of the black dihole that will be considered along the paper, {\it i.e} a configuration of two oppositely charged counterrotating Kerr--Newman (KN) black holes.

%%%%%%%%%%%%%%%%%%%%%%%%%%%%
\section{Equatorial Geodesic Motion}\label{sec3}
%%%%%%%%%%%%%%%%%%%%%%%%%%
In this section, we study equatorial geodesics in the presence of the dihole configuration of two Kerr--Newman black holes. To do so, we start with defining the Lagrangian function ${\cal L}$
as $2{\cal L}=g_{\mu\nu}(dx^{\mu}/d\tau)(dx^{\nu}/d\tau)$. For the choice $z=0$, it can be shown that the metric function $\omega$ vanishes, and the Lagrangian function for the metric (\ref{Papa}) takes the form

\begin{eqnarray}\label{Lag}
2{\cal L}&=&f^{-1}(e^{2\gamma}\dot{\rho}^2+ \rho^2 \dot{\phi}^2)- f\dot{t}^2,
\end{eqnarray}

where the overdot denotes derivation with respect to the affine parameter $\tau$ along the geodesic. The Killing vectors for this spacetime are $\xi_{t}$ and $\xi_{\phi}$, with associate constants of motion $E$ and $L$, respectively, {\it i.e}

\begin{equation}\label{ELe}
\dot{t}=\frac{E}{f},\quad \dot{\phi}= \frac{f L}{\rho^{2}}.
\end{equation}

It should be noted that, in the case of massive particles $E$ is the energy at infinity per mass unit and $L$ is the angular momentum per mass unit, while for massless particles, $E$ is the energy at infinity and $L$ is the angular momentum.

For test particles there exists a third integral of motion, so that the Lagrangian reads $2{\cal L}=-\delta$, with $\delta=1$ for massive particles and $\delta=0$ for photons. Using the previous relations and some straightforward algebra, Eq. (\ref{Lag}) can be conveniently expressed as

\begin{equation}\label{rhodot}
\frac{\dot{\rho}^2 e^{2\gamma}}{2}+V(\rho)=\frac{E^2}{2},
\end{equation}

where $V(\rho)$ denotes the effective potential,

\begin{equation}\label{poteq}
V(\rho)=\frac{f}{2}\left(\delta+\frac{L^2}{\rho^2}f\right).
\end{equation}

\begin{figure}
\centering
\begin{tabular}{cc}
(a)&(b)\\
\includegraphics[width=4.2cm,angle=0]{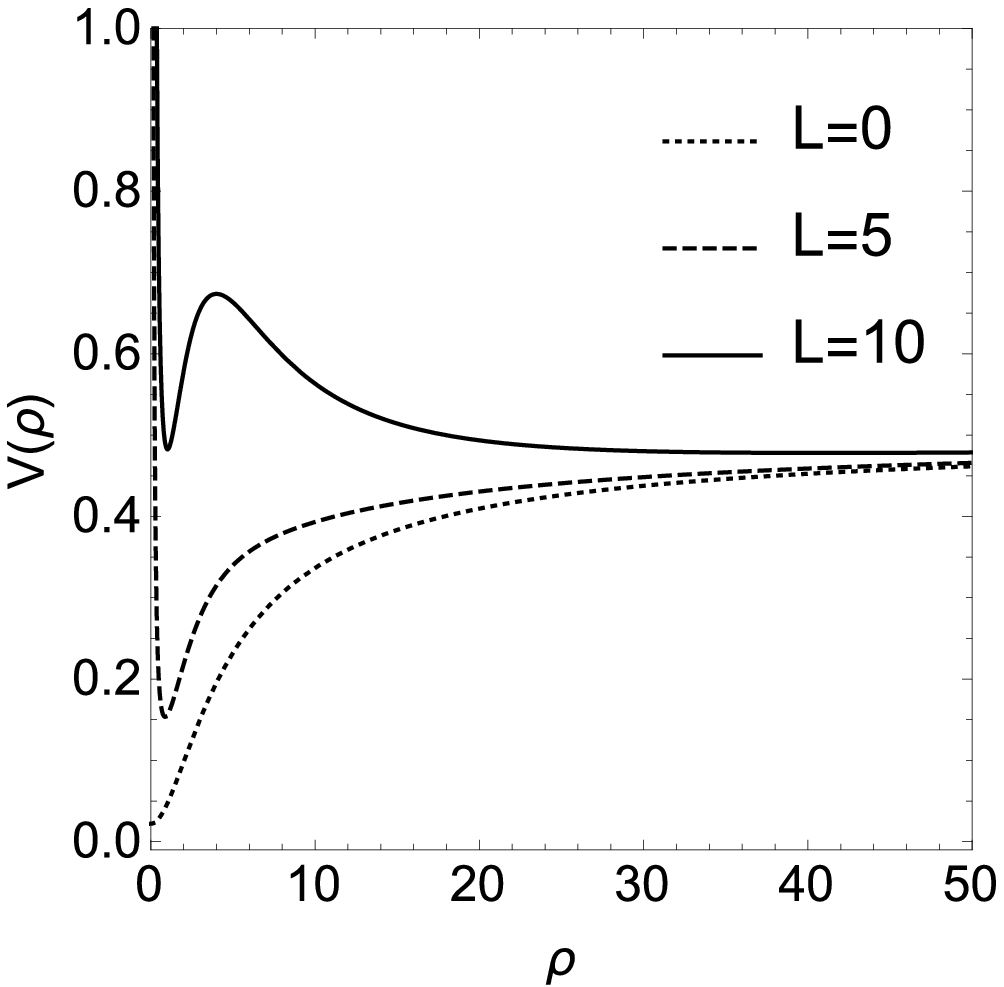}&
\includegraphics[width=4.2cm,angle=0]{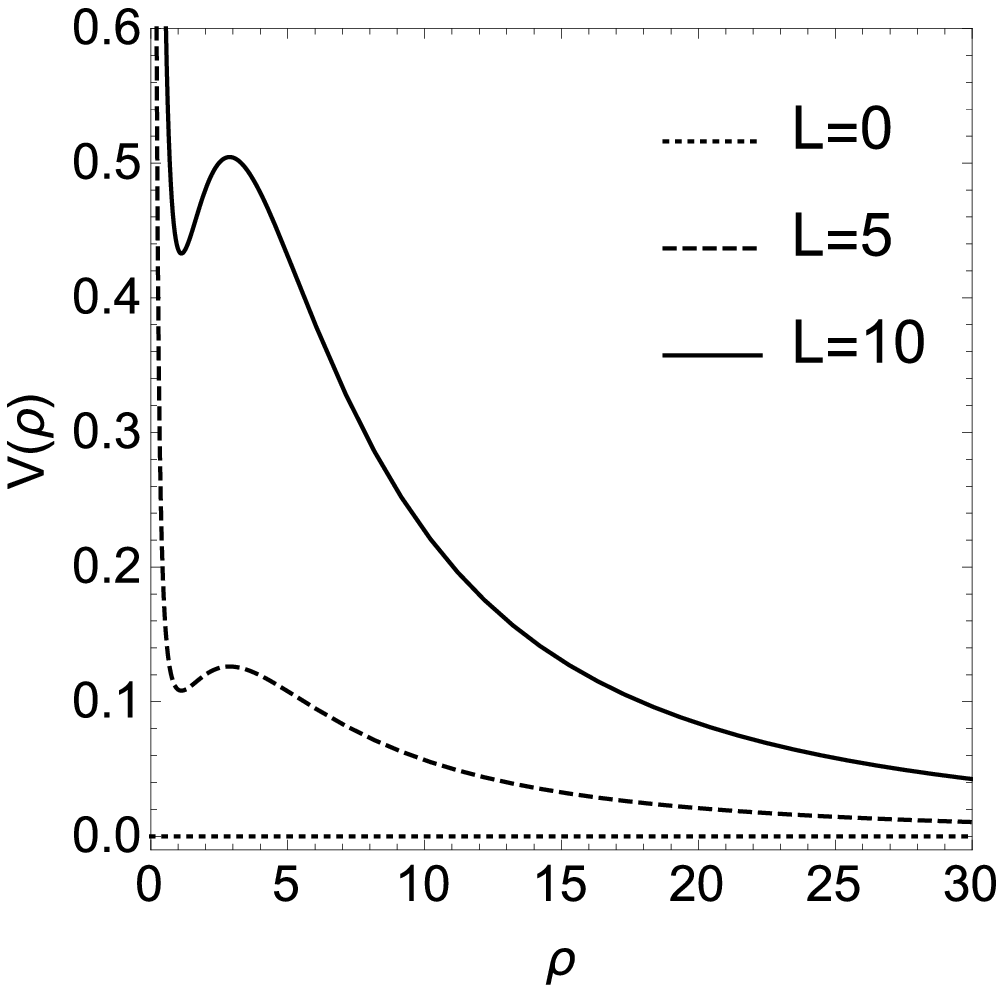}
\end{tabular}
\caption{Effective potential $V(\rho)$ for the dihole solution in the equatorial plane for (a) massive particles and (b) photons, using different values of $L$. The parameters have been set as follows: $M=Q=a=1, R=3$.}
\label{fig2}
\end{figure}

For $L \in \mathbb{R}\,-\,\{0\}$, the effective potential (\ref{poteq}) tends to infinity when $\rho\to 0$. By setting $L=0$, the effective potential for photons vanishes, while the potential for massive particles becomes attractive. Moreover, in the asymptotic limit, $\rho\to\infty$, the potential tends to $\delta/2$ for any $L$. In Figs. \ref{fig2}a and \ref{fig2}b, the effective potential for massive test particles and photons is depicted for different values of angular momentum $L$. 

The corresponding equation of motion for the radial coordinate $\rho$ reads

\begin{eqnarray}\label{eqmoveq}
2f\ddot{\rho}&=&-e^{-2\gamma}[f^2 L^2 \rho^{-3}(\rho f_{,\rho}-2 f)+E^{2}f_{,\rho}]\nonumber\\
&&+\dot{\rho}^2(f_{,\rho}-2 f \gamma_{,\rho}).
\end{eqnarray}

The numerical solution to equation (\ref{eqmoveq}) can be found by defining the parameters $E$, $L$ and the initial condition $\rho_{0}$, which determines $\dot{\rho_{0}}$ through Eq. (\ref{rhodot}).

The possible orbits for time-like and null geodesics can be classified as: (a) circular orbits (stable or unstable), (b) bound precessing orbits and (c) scattered orbits. For equatorial motion, there is no possibility to get plunging orbits due to the position of the constituents out of the equatorial plane ($z=\pm R/2$).  

\begin{figure}
\centering
\begin{tabular}{cc}
(a)&(b)\\
\includegraphics[width=4.0cm,angle=0]{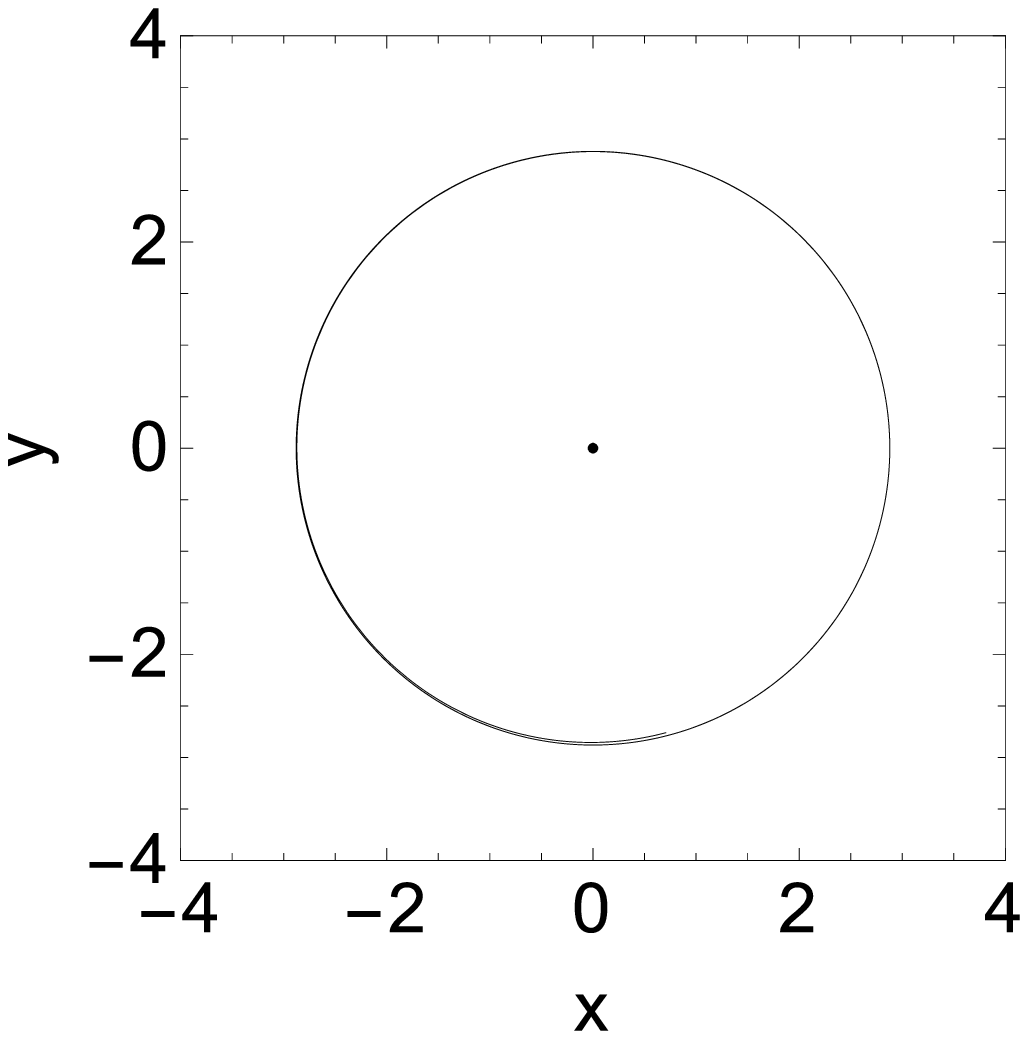}&
\includegraphics[width=4.0cm,angle=0]{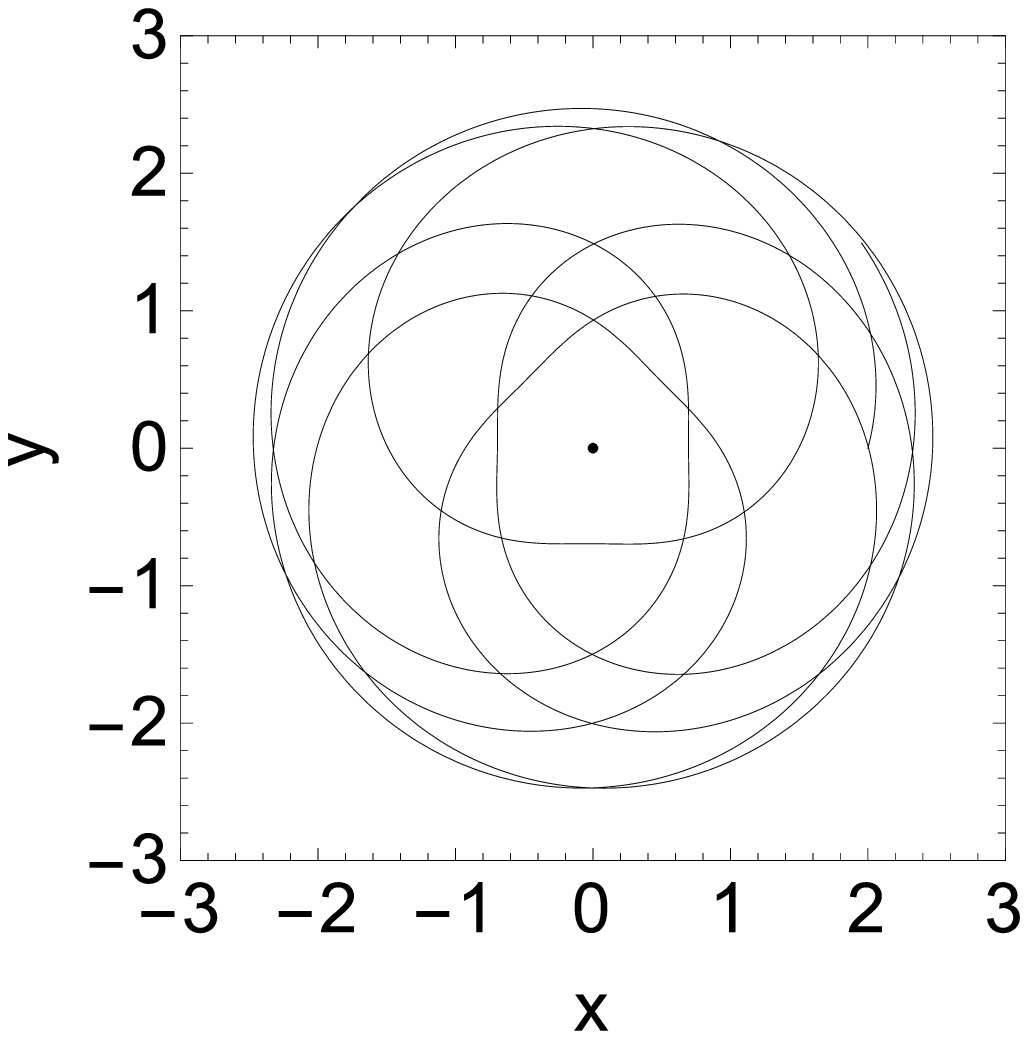}\\
(c)&(d)\\
\includegraphics[width=4.0cm,angle=0]{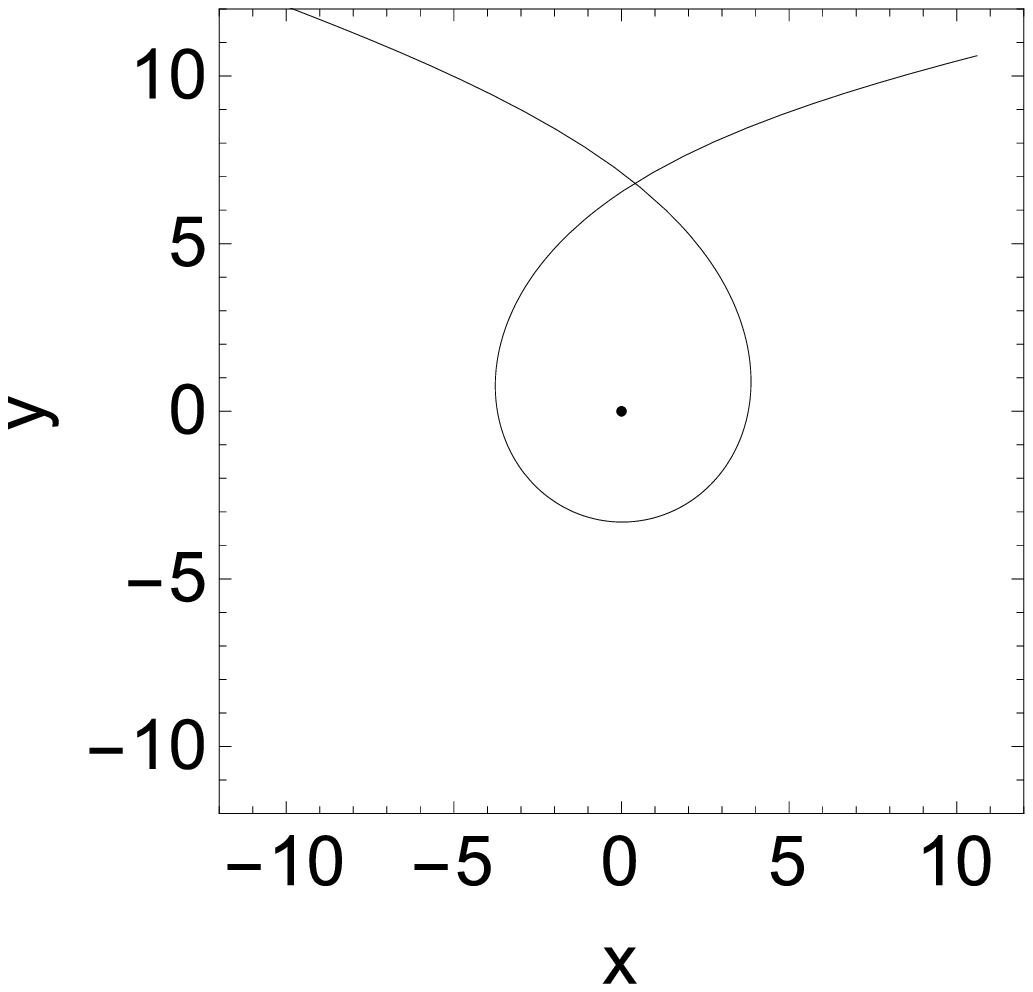}&
\includegraphics[width=4.0cm,angle=0]{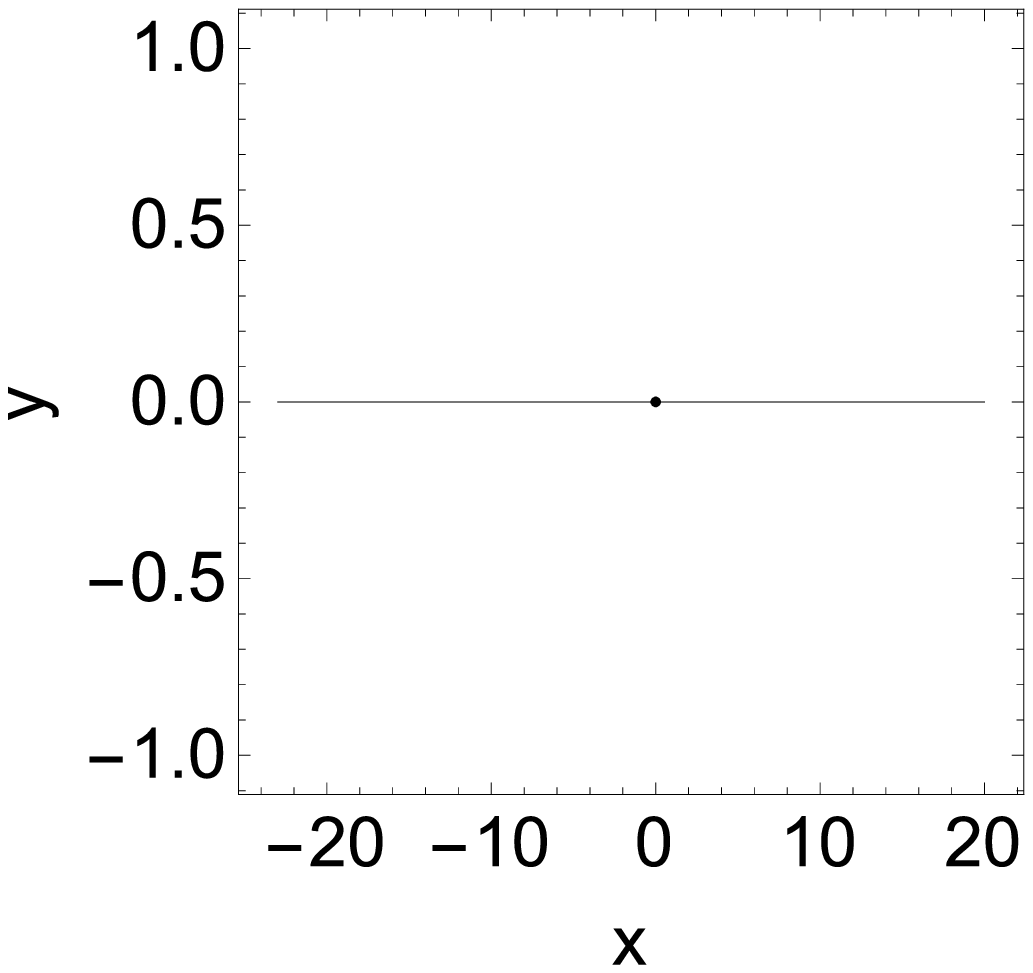}
\end{tabular}
\caption{Different classes of equatorial orbits for photons ($\delta=0$) in a stationary dihole spacetime. The parameters $E$, $L$ and the initial condition $\rho_{0}$ have been set as follows: $E=2.008, L=20$ and $\rho_{0}=2.879$ (a), $E=2, L=20$ and $\rho_{0}=2$ (b), $E=2, L=20$ and $\rho_{0}=15$ (c) and $E=1, L=0$ and $\rho_{0}=20$ (d). The other parameters are $M=Q=a=1, R=3$.} \label{fig3}
\end{figure}

\begin{figure}
\centering
\begin{tabular}{cc}
(a)&(b)\\
\includegraphics[width=4.0cm,angle=0]{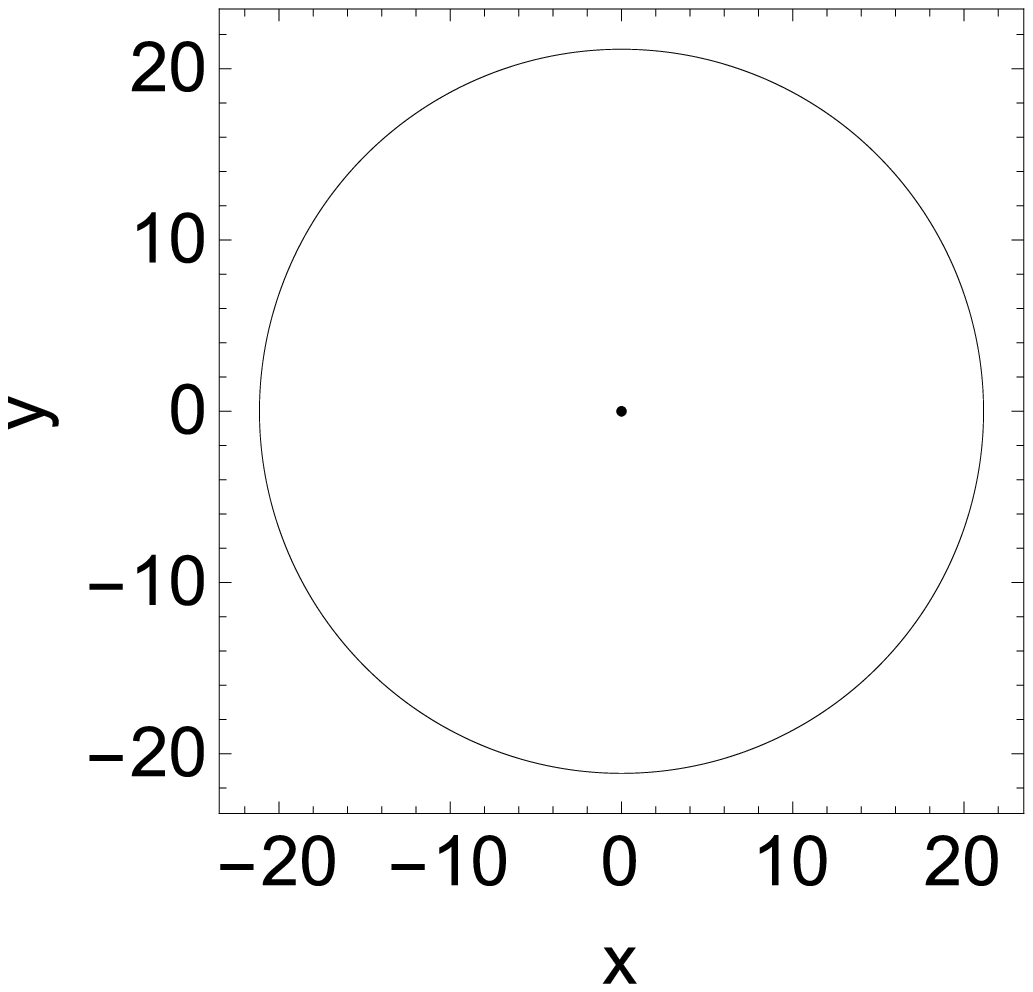}&
\includegraphics[width=4.0cm,angle=0]{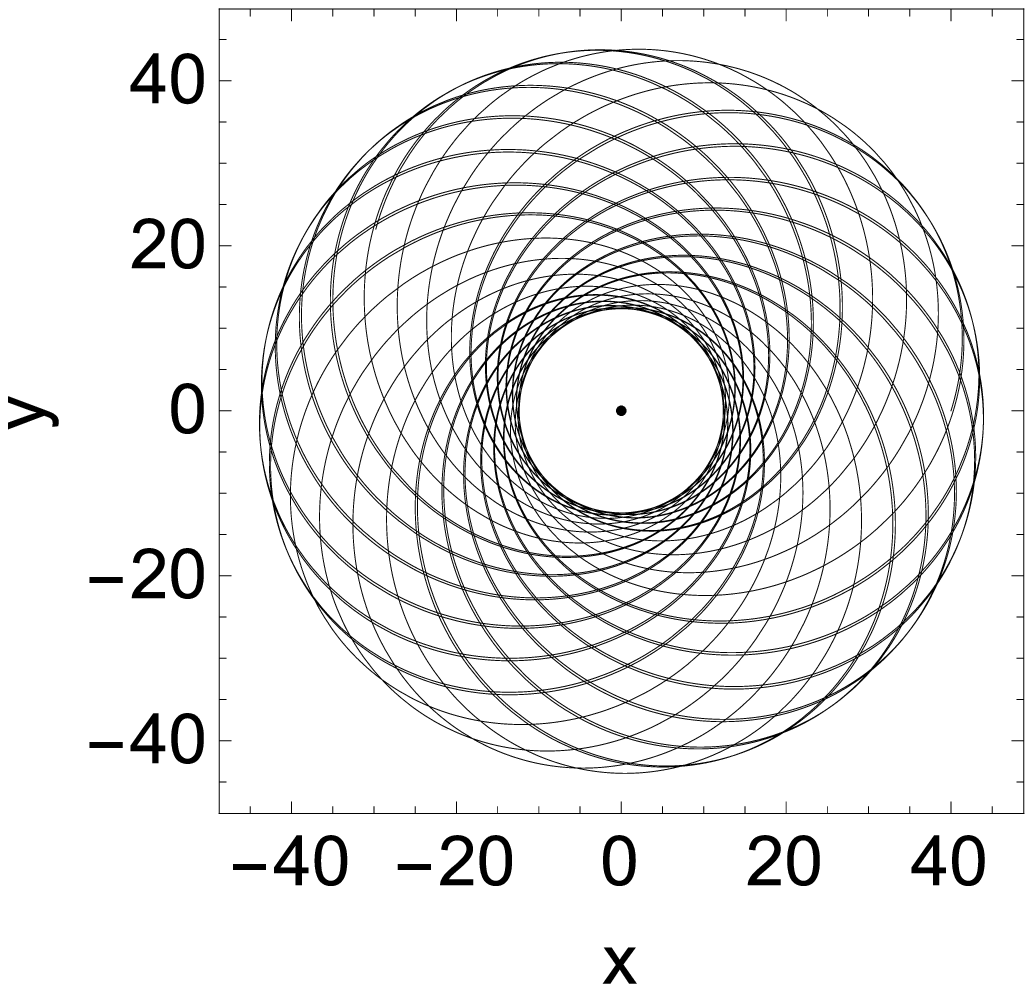}\\
(c)&(d)\\
\includegraphics[width=4.0cm,angle=0]{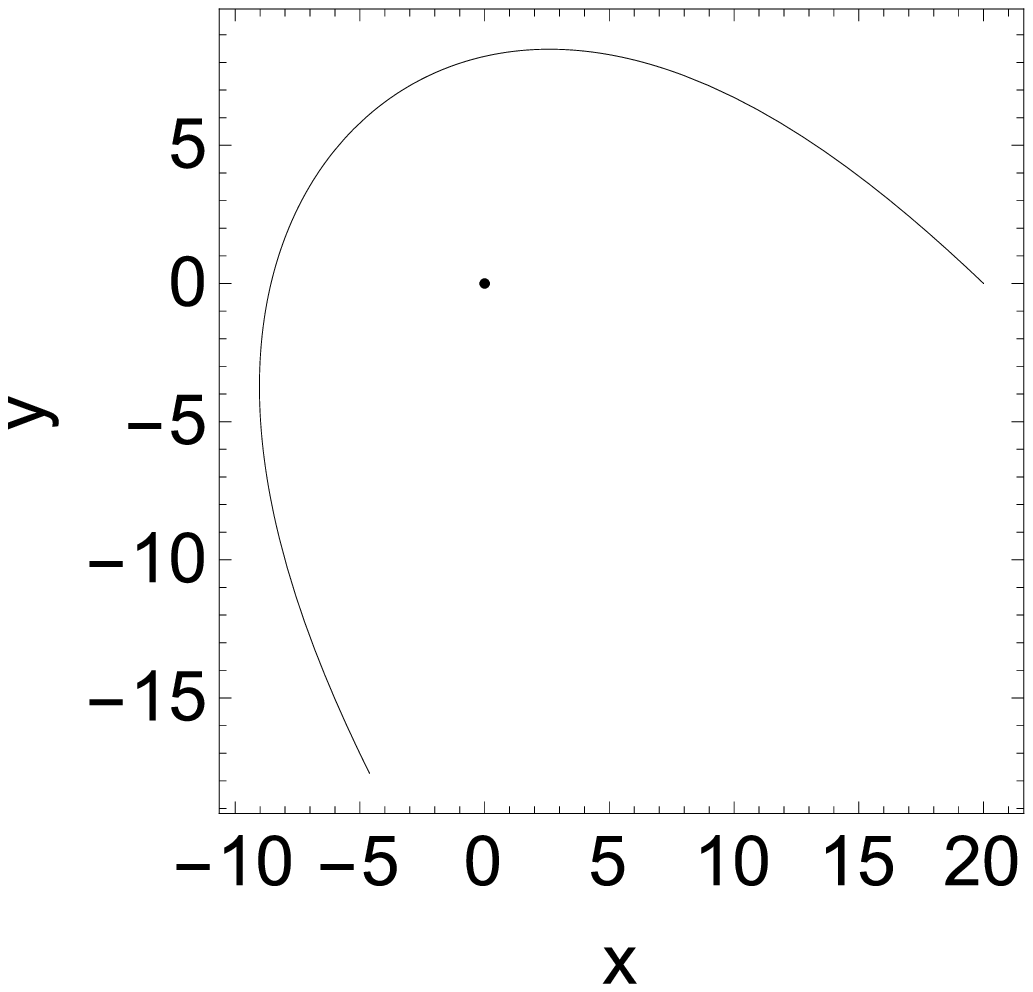}&
\includegraphics[width=4.0cm,angle=0]{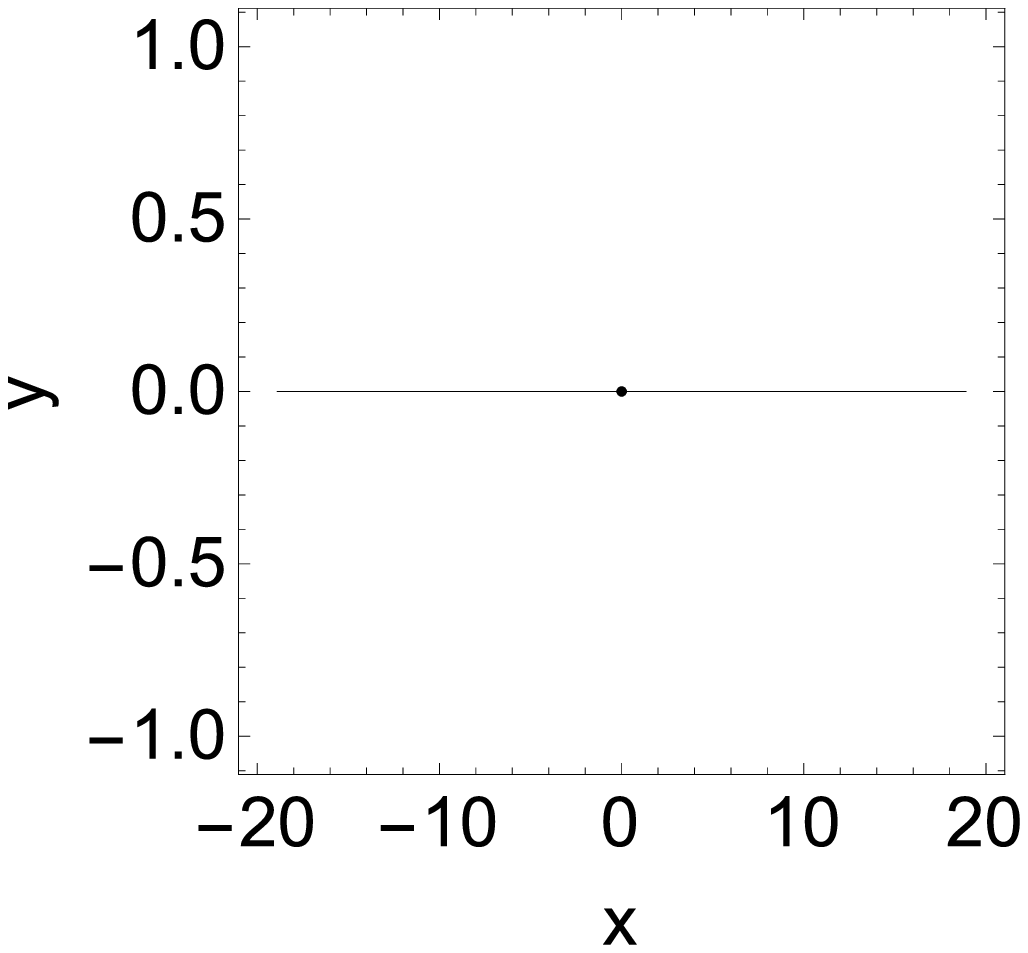}
\end{tabular}
\caption{Different classes of equatorial orbits for massive particles ($\delta=1$) in a stationary dihole spacetime. The parameters $E$, $L$ and the initial condition $\rho_{0}$ have been set as follows: $E=0.961, L=7.9$ and $\rho_{0}=21.142$ (a), $E=0.969, L=7.9$ and $\rho_{0}=40$ (b), $E=1.1, L=10$ and $\rho_{0}=20$ (c) and $E=0.9, L=0$ and $\rho_{0}=10$ (d). The other parameters are $M=Q=a=1, R=3$.} \label{fig4}
\end{figure}

In Figs. \ref{fig3} and \ref{fig4}, we present an example of each type of orbit for photons and massive particles, respectively. Scattered orbits could correspond to unbounded trajectories that after orbiting around the symmetry axis of the system\footnote{In Figs.~\ref{fig3}~and~\ref{fig4}, the central dot at coordinate $(0,0)$ denotes the position of the symmetry axis} escape to infinity (see figs. \ref{fig3}c and \ref{fig4}c), or trajectories that cross the symmetry axis at the same distance from each constituent (see fig. \ref{fig3}d). Circular orbits are also possible for the stationary dihole solution, and can be stable (fig. \ref{fig4}a) or unstable (fig. \ref{fig3}a). The proximity to the symmetry axis of the circular unstable orbits is a consequence of the shape of the effective potential. As can be seen in Figs. \ref{fig2}(a,b), for all $L\neq 0$ there exist a local minimum close to the symmetry axis, which, in the case of photons, and for a given set of parameters $R, M, a$ and $Q$, is always located at the same position. Nonetheless, for massive particles, it is possible to obtain additional local minimum at a larger distance form the symmetry axis.  

\begin{figure}
\centering
\begin{tabular}{cc}
(a)&(b)\\
\includegraphics[height=4.0cm,angle=0]{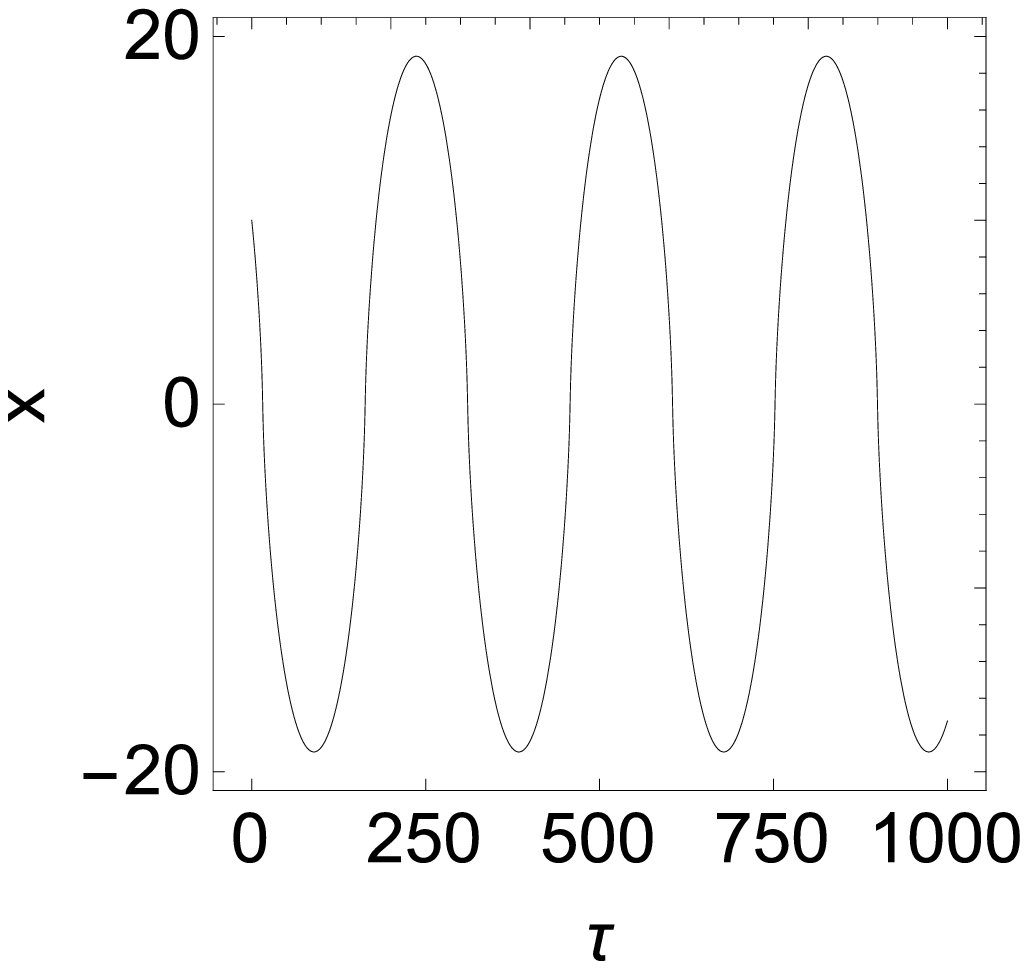}&
\includegraphics[height=4.0cm,angle=0]{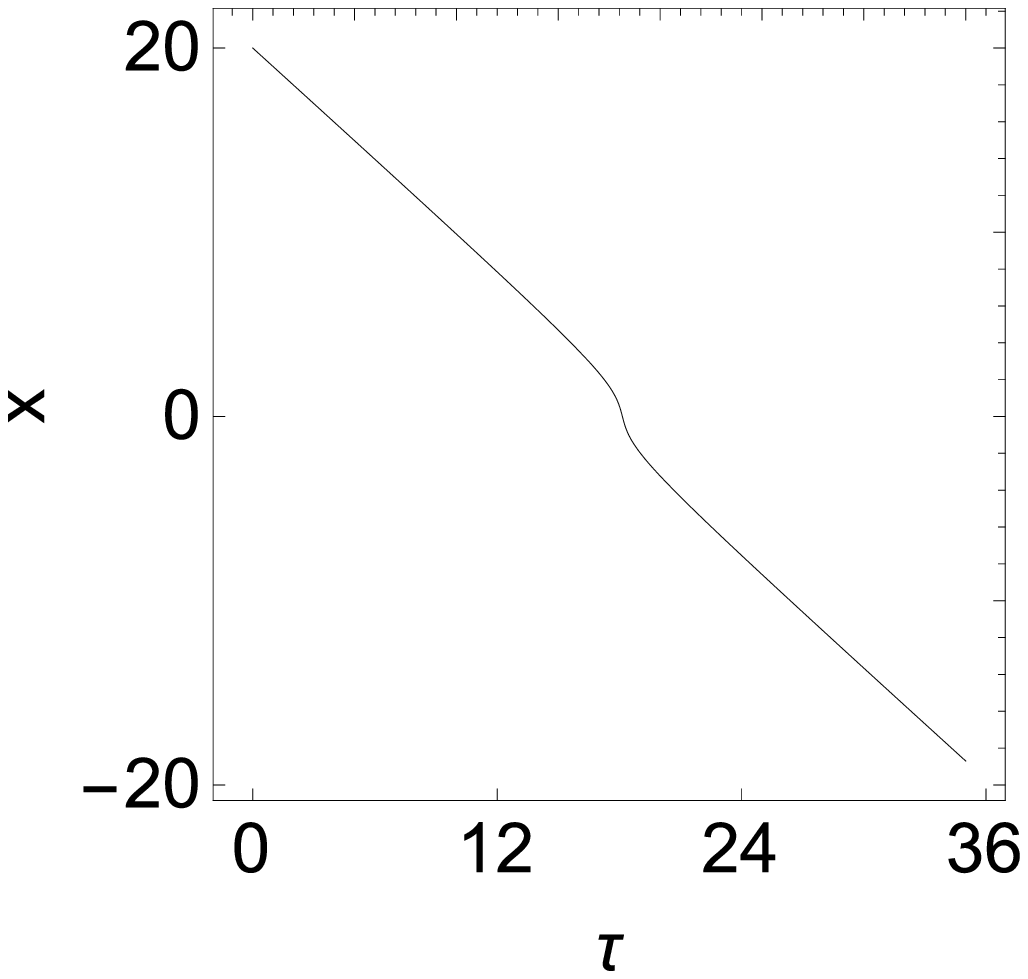}
\end{tabular}
\caption{$x$-position in terms of the affine parameter $\tau$ for (a) massive particles and (b) photons, using the same parameters and initial conditions as in Figs. \ref{fig3}d and \ref{fig4}d.}
\label{fig5}
\end{figure}

For $L=0$, a particular bounded orbit can be reached by massive test particles, which can be considered as the analogue of the Sitnikov problem in the Newtonian restricted three body problem (see Fig.~\ref{fig4}d). Due to the perspective of the figure, it is useful to plot the $x$-position in terms of the affine parameter $\tau$. The results are shown in Fig. ~\ref{fig5}. As can be noted, the trajectory of the massive particle exhibits oscillatory motion \ref{fig5}a, which clearly differs from the unbounded trajectory for photons, Fig. ~\ref{fig5}b. The deviation of the straight line observed at the origin in Fig. ~\ref{fig5}b could obey to the violation of the elementary flatness condition, and can be explained as follows. From Eq. (\ref{rhodot}), photons with $L=0$, satisfy $\dot{\rho}=E e^{-\gamma}$. So, if $\gamma(\rho=0,z=0)\neq 0$ (see Fig. \ref{lgamma}), the derivative of $\rho$ with respect to the affine parameter is not a constant at the origin, giving place to the observed shift. 

Furthermore, in the equatorial plane of the dihole there exist two more classes of bound precessing orbits (Figs. \ref{fig3}b and \ref{fig4}b). In the first case, the orbit is formed by an internal repulsion close to the symmetry axis, whereas in the second case the test particle follows a typical orbit produced by an attractive potential. It should be pointed out that both kinds of orbits might appear for massive particles. However, when considering photons, the bounded precessing orbit always exhibits repulsion in the inner radius, as expected from the effective potential Eq. (\ref{poteq}).

The radii of stable circular orbits have a minimum value named innermost stable circular orbit.  Each such orbit can be determined as follows \cite{sanabria}: {\it i)} Define the effective potential as a function of the parameters $E$ and $L$; {\it ii)} solve $V(\rho; E, L)=0$ and $d V(\rho; E, L)/d\rho=0$ for $E$ and $L$; {\it iii)} calculate the second derivative $d^2 V(\rho)/d\rho^2$ and replace the previous values of $E$ and $L$ ; {\it iv)} solve $d^2 V(\rho)/d\rho^2=0$ for $\rho$ and {\it v)} determine the circumferential ISCO-radius as $R_{{\rm ISCO}}=\sqrt{g_{\phi\phi}}$.

Redefining the effective potential in terms of $E$ and $L$ as

\begin{equation}
V(\rho; E, L)=E^2-\frac{f^2 L^2}{\rho^2}-\delta f,
\end{equation}

and following steps {\it ii)} and {\it iii)} described above,  the second derivative of the potential can be written as 

\begin{eqnarray}
\frac{d^2 V(\rho)}{d\rho^2}&=&\delta  \bigg\{f^2 \bigg[\rho ^2 f_{,\rho\rho\rho} f_{,\rho}-2 \big(f_{,\rho}-\rho f_{,\rho \rho}\big)^2\bigg]-f^3 \nonumber\\
&\times&\big(\rho f_{,\rho \rho\rho} +4 f_{,\rho \rho}\big)+\rho^2 (f_{,\rho})^3  \big(\rho f_{,\rho\rho} -2 f_{,\rho}\big)\nonumber\\
&+&f \rho (f_{,\rho})^2 (4 f_{,\rho}-\rho f_{,\rho \rho})\bigg\}.
\end{eqnarray}

The last third order differential equation gives a trivial solution for photons, since $\delta=0$. So, in what follows, we restrict ourselves to the study of massive particles. 

\begin{figure}
\centering
\begin{tabular}{cc}
\hspace{0.8cm}(a)&\hspace{0.8cm}(b)\\
\includegraphics[width=4.0cm,angle=0]{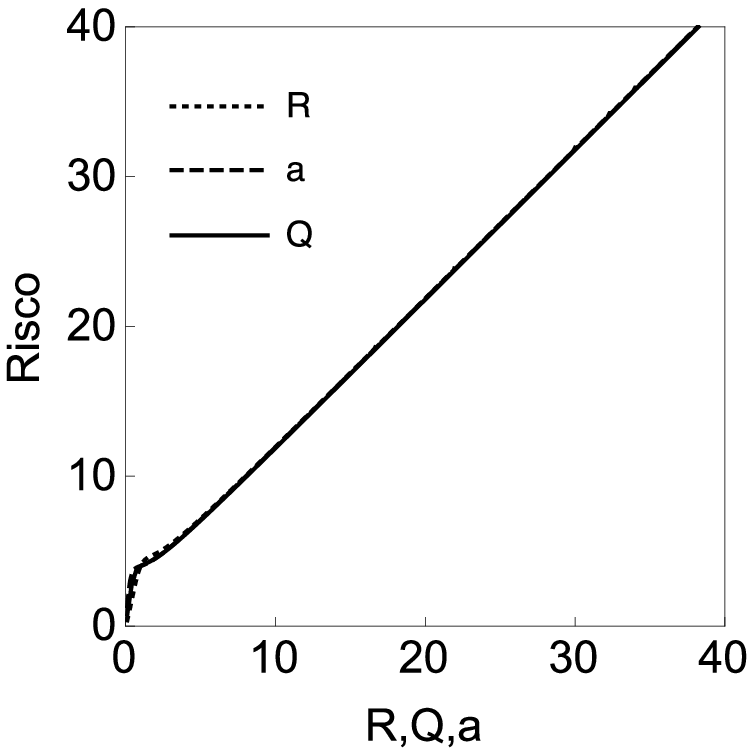}&
\includegraphics[width=4.0cm,angle=0]{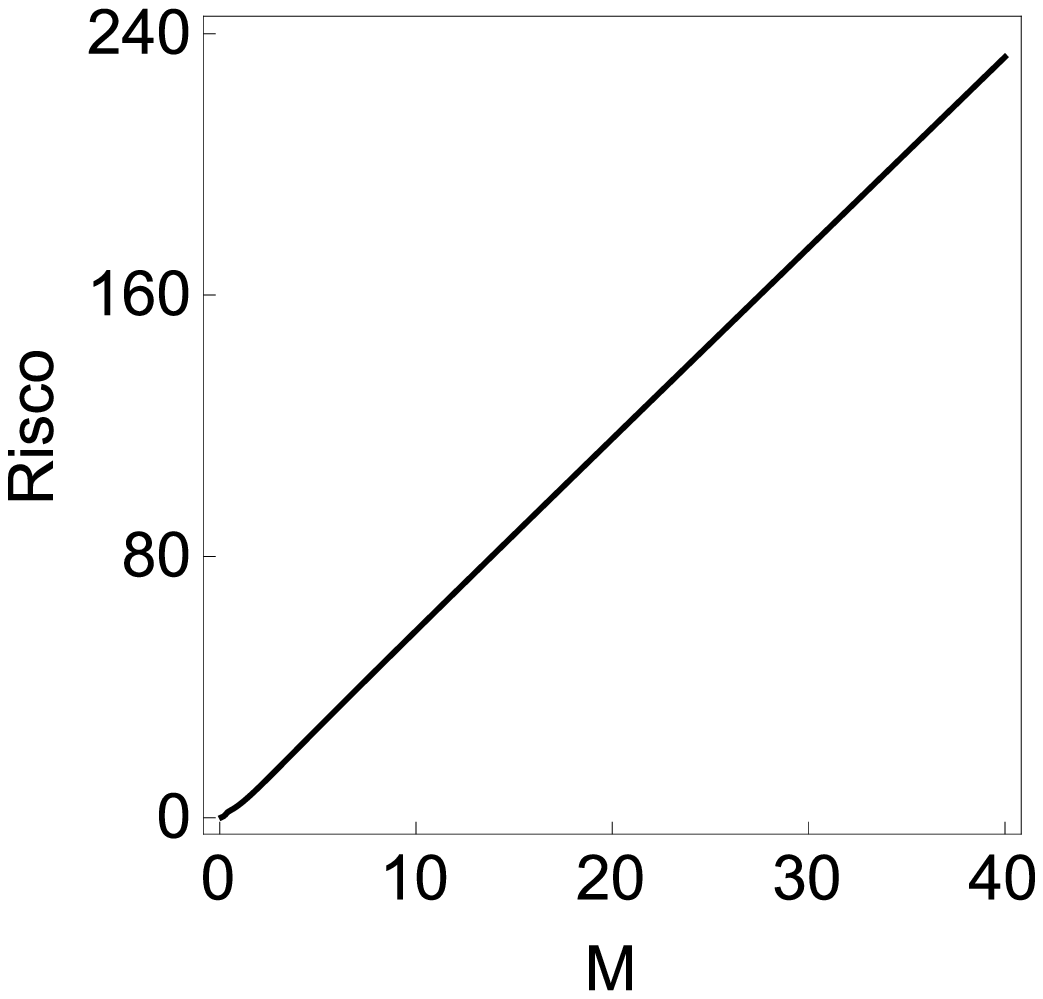}
\end{tabular}
\caption{Innermost stable circular orbits $R_{{\rm ISCO}}$ for the dihole solution in terms of  $Q$ (continuous line), $R$ (dotted line) and $a$ (dashed line) (left panel), and in terms of $M$ (right panel). In each case, the remaining parameters are set to 1.} 
\label{fig6}
\end{figure}

In Fig.~\ref{fig6}, we present the $R_{{\rm ISCO}}$ as a function of the physical parameters of the source, $R, M, a$ and $Q$. As expected in the limit $M\to 0$ the $R_{{\rm ISCO}}$ tends to zero (Fig. \ref{fig6}b). Another interesting result from this figure, is that the curves for $R_{{\rm ISCO}}$ as a function of $M, Q, R$ and $a$, deviate slightly away from straight lines (with a much less pronounced effect for $M$), for small values of the physical parameters. Furthermore, the global dependence of the radius at ISCO with the parameters $R, Q$ and $a$, exhibits a very similar behavior (Fig. \ref{fig6}a), with an approximate slope of 1.07, while in Fig. \ref{fig6}b the slope is approximately 5.81. A superposition of Figs.~\ref{fig6}(a)~and~\ref{fig6}(b), shows a crossing point at $R_{{\rm ISCO}}\approx 4.01$, corresponding to the case $R=M=Q=a=1$, with radial Weyl coordinate $\rho \approx 1.03$.

%%%%%%%%%%%%%%%%%%%%%%%%%%%%
\section{Non-equatorial Geodesic Motion}\label{sec4}
%%%%%%%%%%%%%%%%%%%%%%%%%%
Let us consider the more general problem of a test particle moving with no restriction to the equatorial plane. The corresponding Lagrangian is given by

\begin{eqnarray}\label{Lag2}
2{\cal L}&=&f^{-1}\left[e^{2\gamma}\left(\dot{\rho}^2 + \dot{z}^2\right)+ \rho^2 \dot{\phi}^2\right]- f(\dot{t}-\omega \dot{\phi})^2.
\end{eqnarray}

From the cyclic coordinates $t$ and $\phi$, we get

\begin{equation}\label{EL}
\dot{t}=\frac{E}{f}+\frac{f\omega(L-E \omega)}{\rho^{2}},\quad \dot{\phi}= \frac{f(L-E \omega)}{\rho^{2}},
\end{equation}

where the constants $E$ and $L$, are related to the Killing vectors $\xi_{t}$ and $\xi_{\phi}$ respectively, and have the same meaning as defined for the equatorial case. Following a similar procedure as in section \ref{sec3}, replacing equations (\ref{EL}) and using $2{\cal L}=-\delta$, the effective potential can be defined as

\begin{equation}\label{potef}
V(\rho,z)=e^{-2\gamma } \left(E^2-\frac{f^2 (L-E\omega)^2}{\rho ^2}-f\delta\right).
\end{equation}

By the definition of $V(\rho,z)$, the motion must be restricted to the region $V\geq 0$.

Using the Lagrangian formalism, the equations of motion for the test particles assume the form

\begin{eqnarray}\label{Ecmovrho}
\ddot{\rho}&=&\frac{f e^{-2\gamma}(L-E\omega)}{2\rho^{2}}\left[2 f E \omega_{,\rho}-\left(f_{,\rho}-\frac{2 f}{\rho}\right)(L-E\omega)\right]\nonumber\\
&&-\frac{\dot{\rho}\dot{z}}{f}(2 f \gamma_{,z}-f_{,z})-\frac{(\dot{\rho}^{2}-\dot{z}^{2})}{2f}(2 f \gamma_{,\rho}-f_{,\rho})\nonumber\\
&&-\frac{E^2 f_{,\rho} e^{-2\gamma}}{2f},\\
\ddot{z}&=&\frac{f e^{-2\gamma}(L-E\omega)}{2\rho^{2}}\left[2 f E \omega_{,z}-f_{,z}(L-E\omega)\right]\nonumber\label{Ecmovz}\\
&&-\frac{\dot{\rho}\dot{z}}{f}(2 f \gamma_{,\rho}-f_{,\rho})+\frac{(\dot{\rho}^{2}-\dot{z}^{2})}{2f}(2 f \gamma_{,z}-f_{,z})\nonumber\\
&&-\frac{E^{2} f_{,z}e^{-2\gamma}}{2 f}.
\end{eqnarray}

The system of equations (\ref{Ecmovrho}-\ref{Ecmovz}) was integrated using a Runge-Kutta-Fehlberg (4-5) algorithm with adaptive step size. Given the constants $E$, $L$ and the initial conditions $\rho(0)$, $z(0)$, $p_{\rho}(0)$, the third integral of motion $2{\cal L}=-\delta$ determines $p_{z}(0)$. In Fig.~\ref{fig7}, we present the orbits for timelike geodesics. It is worth noting that the non-equatorial geodesics behave in a very similar way to the orbits in the equatorial case, and can be classified as circular (Figs.~\ref{fig7}(a,b)), bounded (precessing Figs.~\ref{fig7}(c,d) or oscillating Figs.~\ref{fig7}(e,f)), and plunging orbits (Figs.~\ref{fig7}(g,h)). A main difference with the equatorial motion is the existence of plunging orbits. By virtue of the equatorial antisymmetry of the solution, the effective potential is not symmetric under reflections about the plane $z=0$ (which can be seen clearly in Fig.~\ref{fig7}(g) and Fig.\ref{fig8}(e)). It is important to emphasize that there exists a repulsive potential between the constituents which do not let the particles to fall into them through the $z$-axis. In Figs.~\ref{fig7}(g,h), we show the trajectories followed by massive particles with eight different initial conditions. By setting $L=0$, the repulsive potential between the dihole disappears and the oscillating solutions take place.

On the other hand, in Fig.~\ref{fig8} the orbits for null geodesics are presented. As in the case of timelike geodesics, for $L\neq 0$, the section $(-R/2+\sigma, R/2-\sigma)$ cannot be reached by the photons (see figs. \ref{fig8}(c,d,e,f)), while for $L=0$ the repulsive potential disappears and the photon may cross between the sources as in Fig.~\ref{fig5}b (see Figs.~\ref{fig8}(a,b,g,h)). 

%%%%%%%%%%%%%%%%%%%%%%%%%%%%
\section{Concluding Remarks}
\label{Conclusions}%%%%%

In this paper, we have studied timelike and null geodesics in the presence of a dihole configuration of two counter--rotating Kerr--Newman black holes with equal and opposite charges.  For geodesics in the equatorial plane, it was possible to classify the orbits into three different types: circular, precessing and scattered orbits. Plunging orbits do not exist, due to the position of the dihole constituents out of the equatorial plane at $z=\pm R/2$. As a result of this study, we found that close to the symmetry axis, some bounded precessing orbits for massive particles and photons exhibit repulsion in the inner radii.  We also showed that when the angular momentum of the test particle tends to zero, the repulsive potential between the sources vanishes. Under this conditions, a photon traveling at an equidistant position from the constituents, and in absence of repulsion, experiences a shift close to the dihole, while the trajectory of a massive particle experiences an oscillatory motion (as a first analog of the Sitnikov problem in General Relativity). We conjecture that the deviation in the trajectory of null geodesics is due to the lack of elementary flatness in the interval $(-R/2+\sigma, R/2-\sigma)$.

Many interesting phenomena can happen in the innermost stable circular orbit of the black dihole.  For example, we found that the $R_{\rm ISCO}$ satisfies a linear relation with the mass of the constituents $M$. The slope $m$ of the graph $R_{\rm ISCO}$ vs. $M$, slightly deviates from the radius for the Schwarzschild metric, and takes the value $m \approx 5.8$ (right panel in Fig. \ref{fig6}). This effect could be explained in terms of the contributions of the multipolar moments of the solution (see {\it e.g.} \cite{Shibata1998} and \cite{sanabria}). On the other hand, when considering the dependence of the radius of the ISCO with the angular momentum $J$, electric charge $Q$ and separation between the sources $R$, a similar (almost linear) dependence with the three parameters is observed. Nevertheless, for numerical values of the physical parameters $R, Q$ and $a$ below 3, the dependence between the radius of the ISCO and the parameters is nonlinear. These results, suggest a dominant role of the linear terms relating the ISCO-radius with the physical parameters of the dihole.  

Finally, we have solved the equations of motion for non-equatorial timelike and null geodesics. Similar results to the ones in the restricted equatorial motion were found, along with the fact that plunging orbits are possible. However, the plunging orbits differ with previous results found by Chandrasekhar for null geodesics in the field of two static black holes \cite{Chandrasekhar1989}, that is, in the stationary dihole solution the test particles do not turn around one of the constituents before being trapped within the black holes, as it does in the case of two extreme Reissner-Nordstr{\"o}m black holes. The differences are associated to the existence of the strut, being that the Reissner-Nordstr{\"o}m dihole reaches equilibrium by means of the balance between their electrostatic repulsion and the gravitational attraction.
%%%%%%%%%%%%%%%%%%%%%%%%%%
\section*{Acknowledgments}
%%%%%%%%%%%%%%%%%%%%%%%%%%
We thank Professor Vladimir S. Manko for originally directing our attention to study geodesic motion around the dihole solution and subsequent discussions and comments at the early stages of this paper. We also like to thank F. D. Lora-Clavijo for reading an earlier draft of the present paper, and valuable critical comments. FLD acknowledges financial support from the University of the Llanos, under Grants Commission: Postdoctoral Fellowship Scheme.

%%%%%%%%%%%%%%%%%%%%%%%%%%%

%\newpage

\begin{widetext}

\begin{figure}
\centering
\begin{tabular}{cccc}
(a)&(b)&(c)&(d)\\
\includegraphics[width=4cm,angle=0]{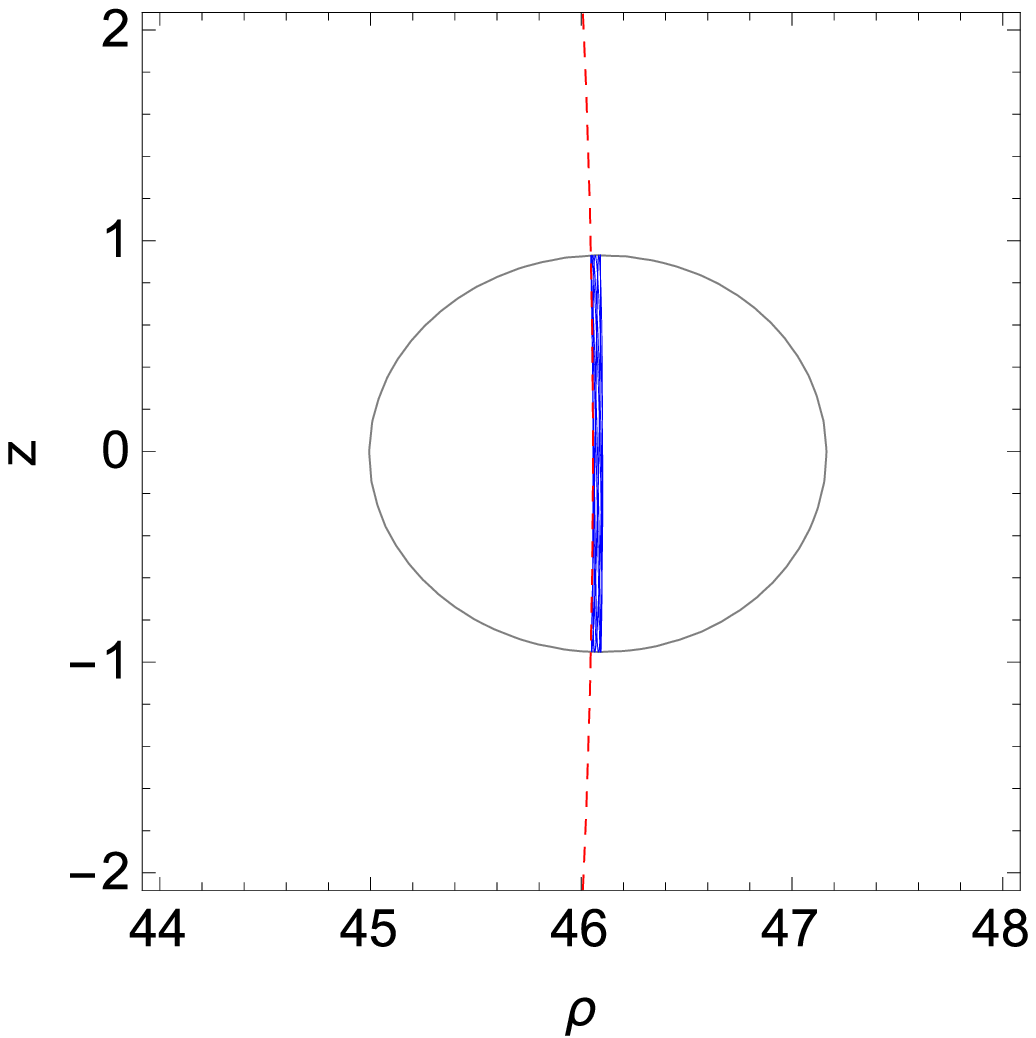}&
\includegraphics[width=4cm,angle=0]{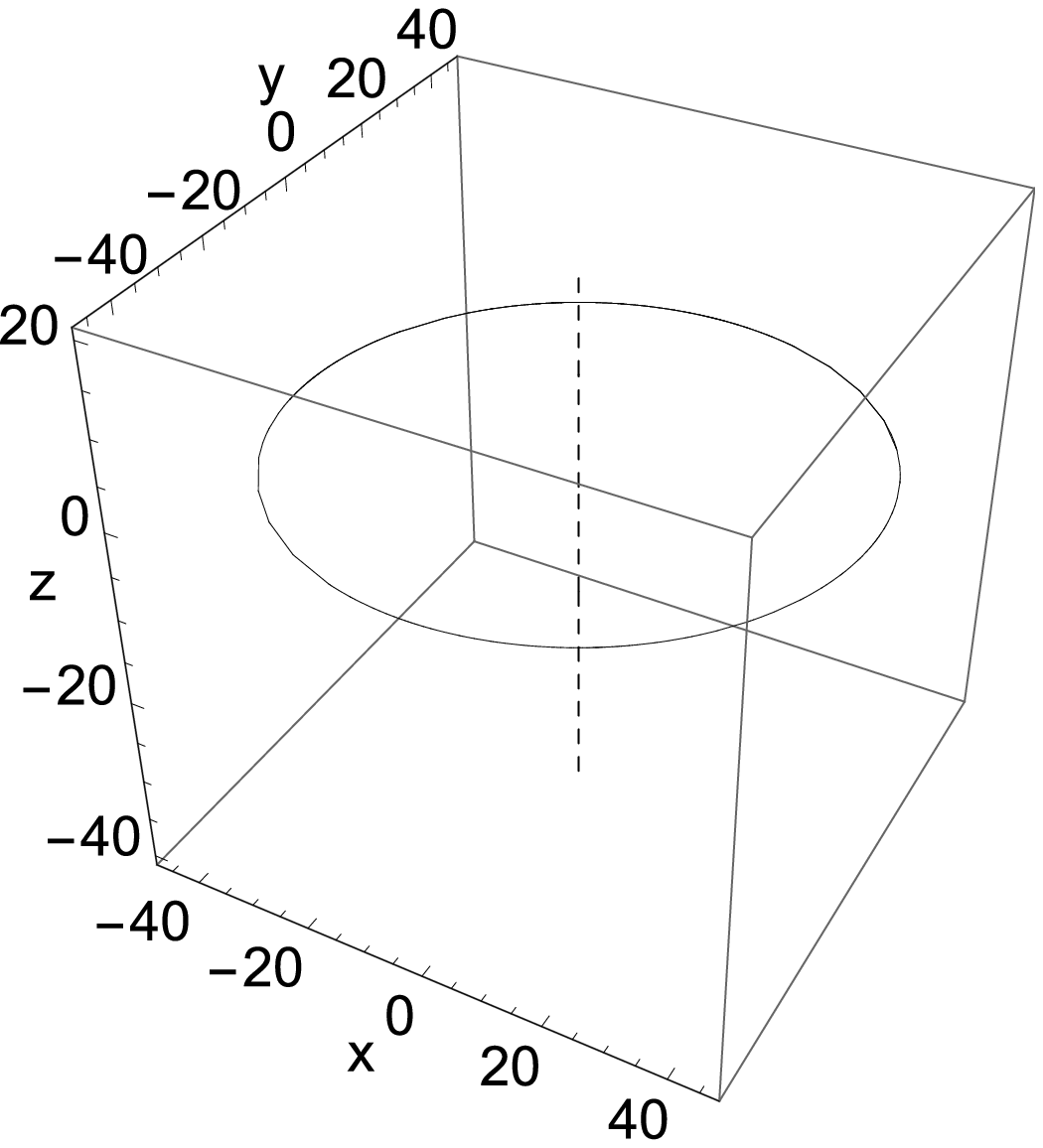}&
\includegraphics[width=4cm,angle=0]{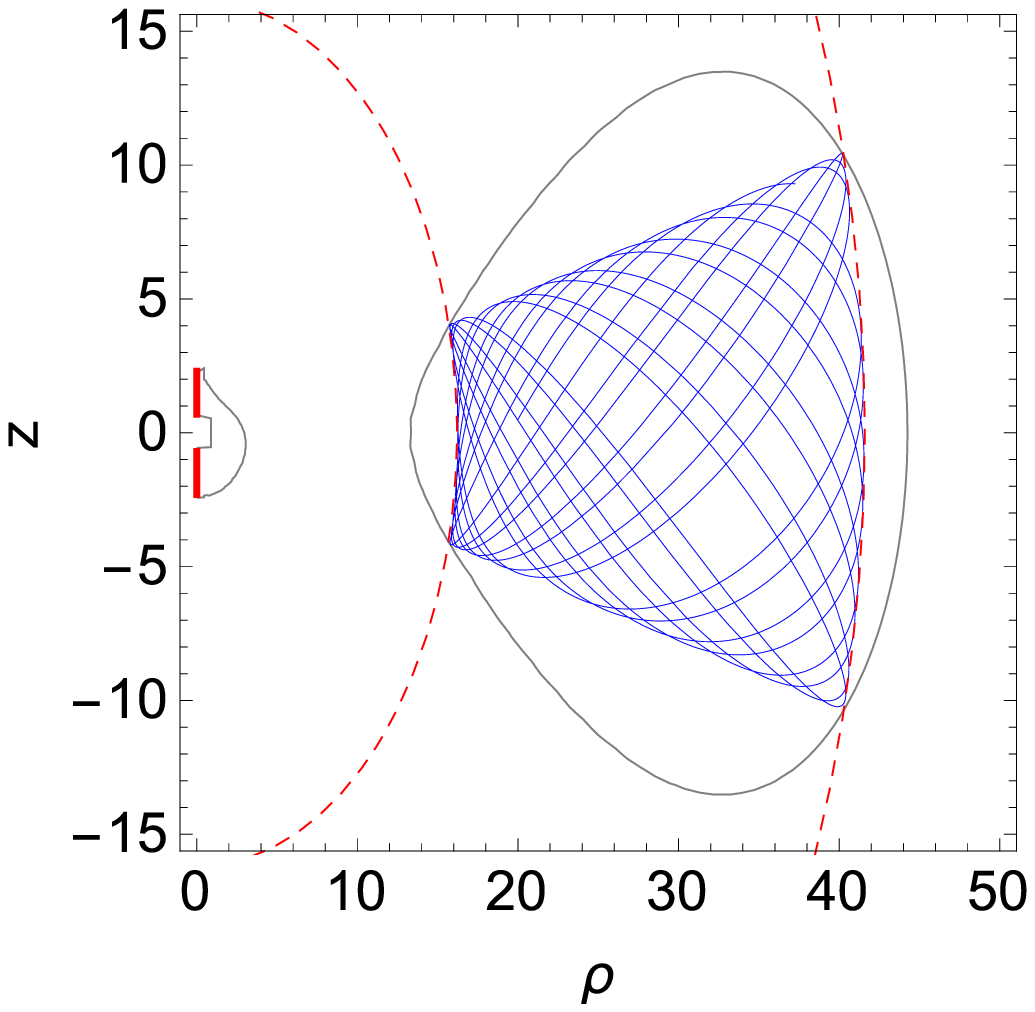}&
\includegraphics[width=4cm,angle=0]{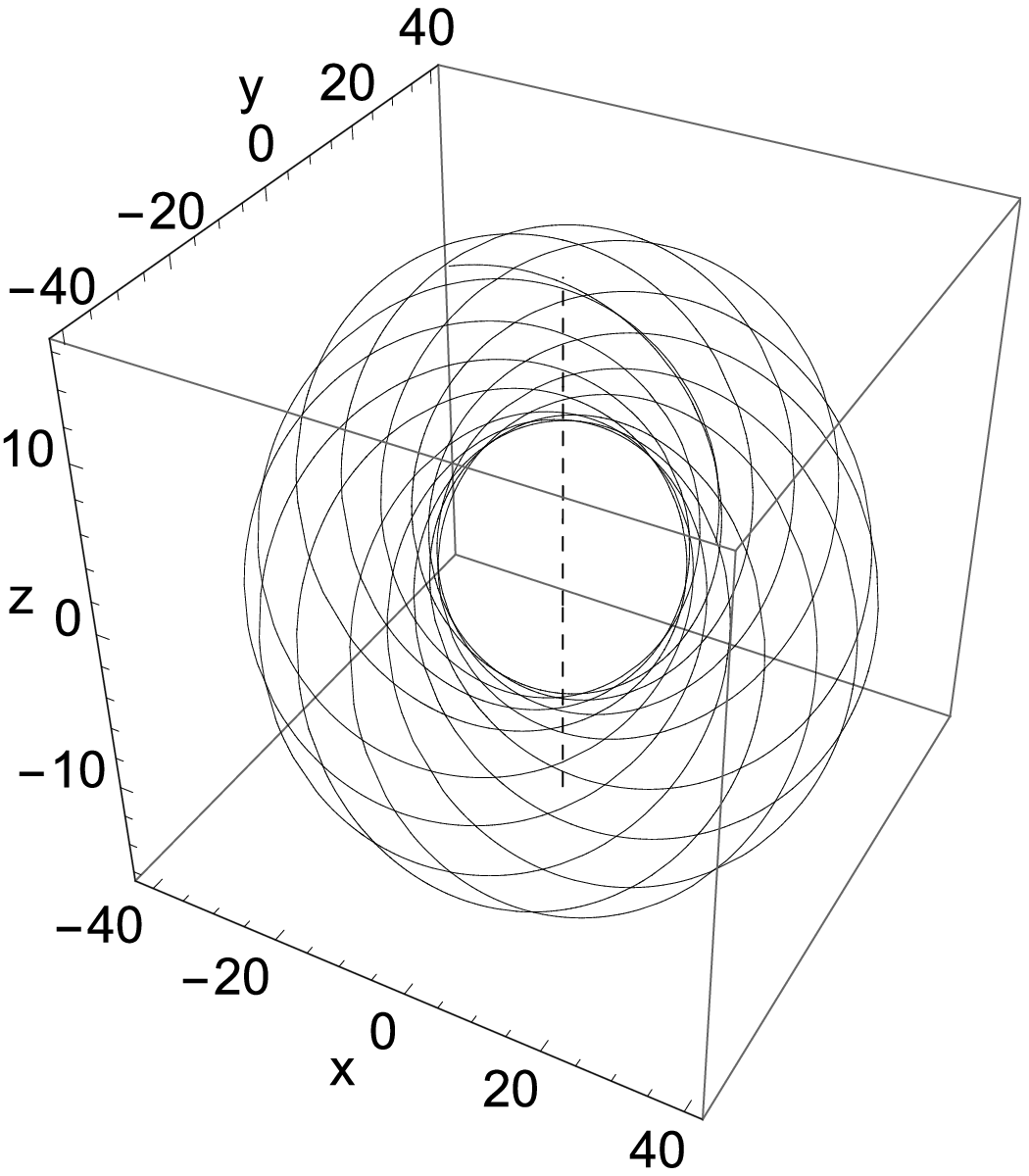}\\
(e)&(f)&(g)&(h)\\
\includegraphics[width=4cm,angle=0]{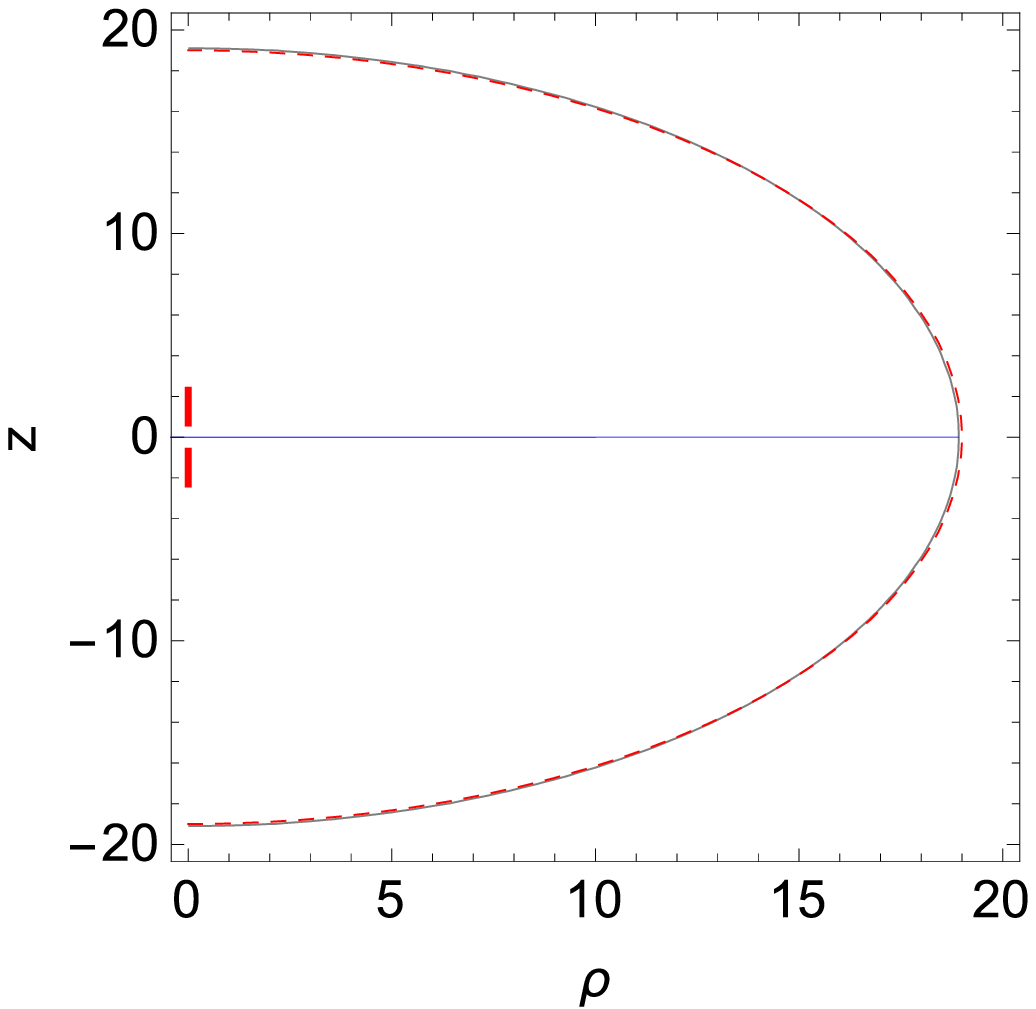}&
\includegraphics[width=4cm,angle=0]{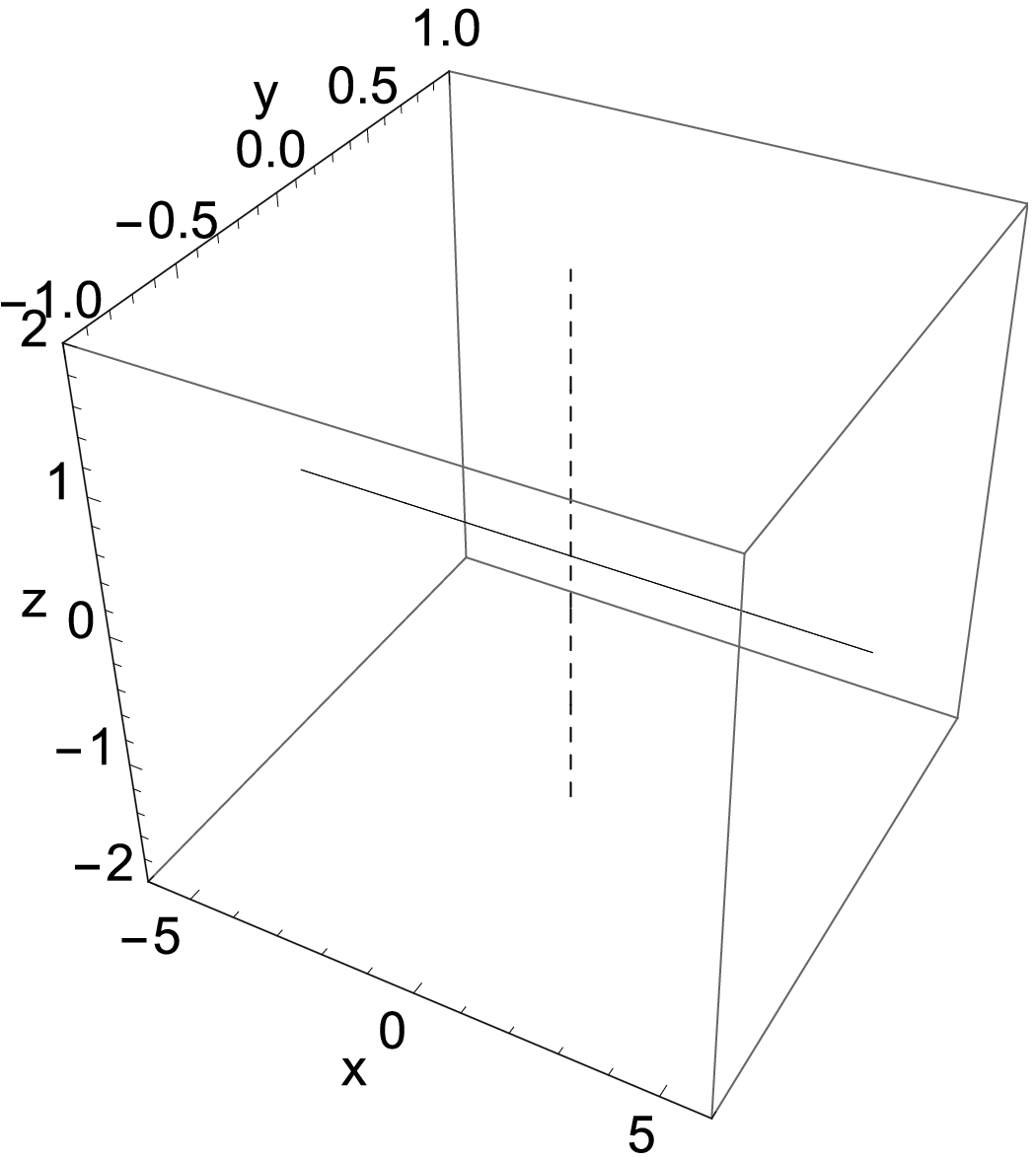}&
\includegraphics[width=4cm,angle=0]{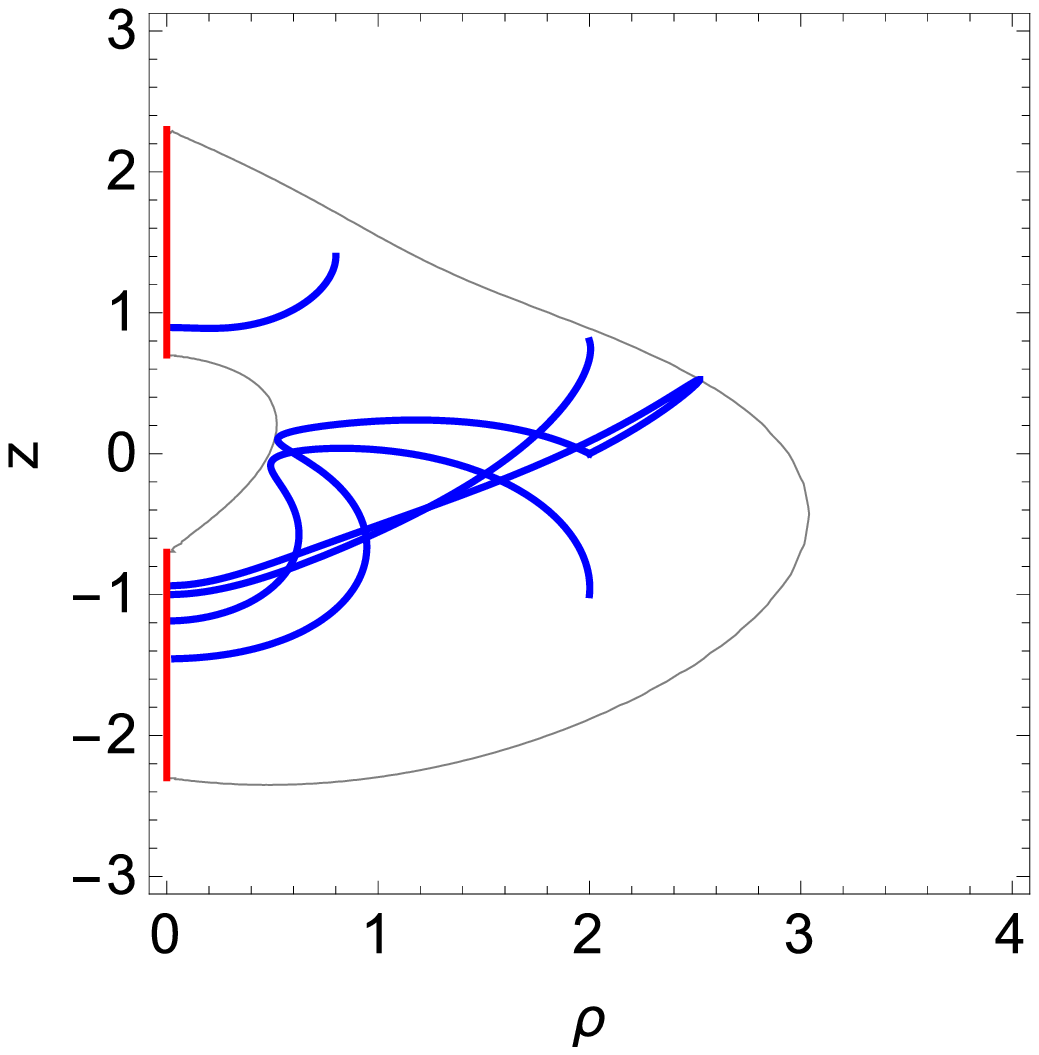}&
\includegraphics[width=4cm,angle=0]{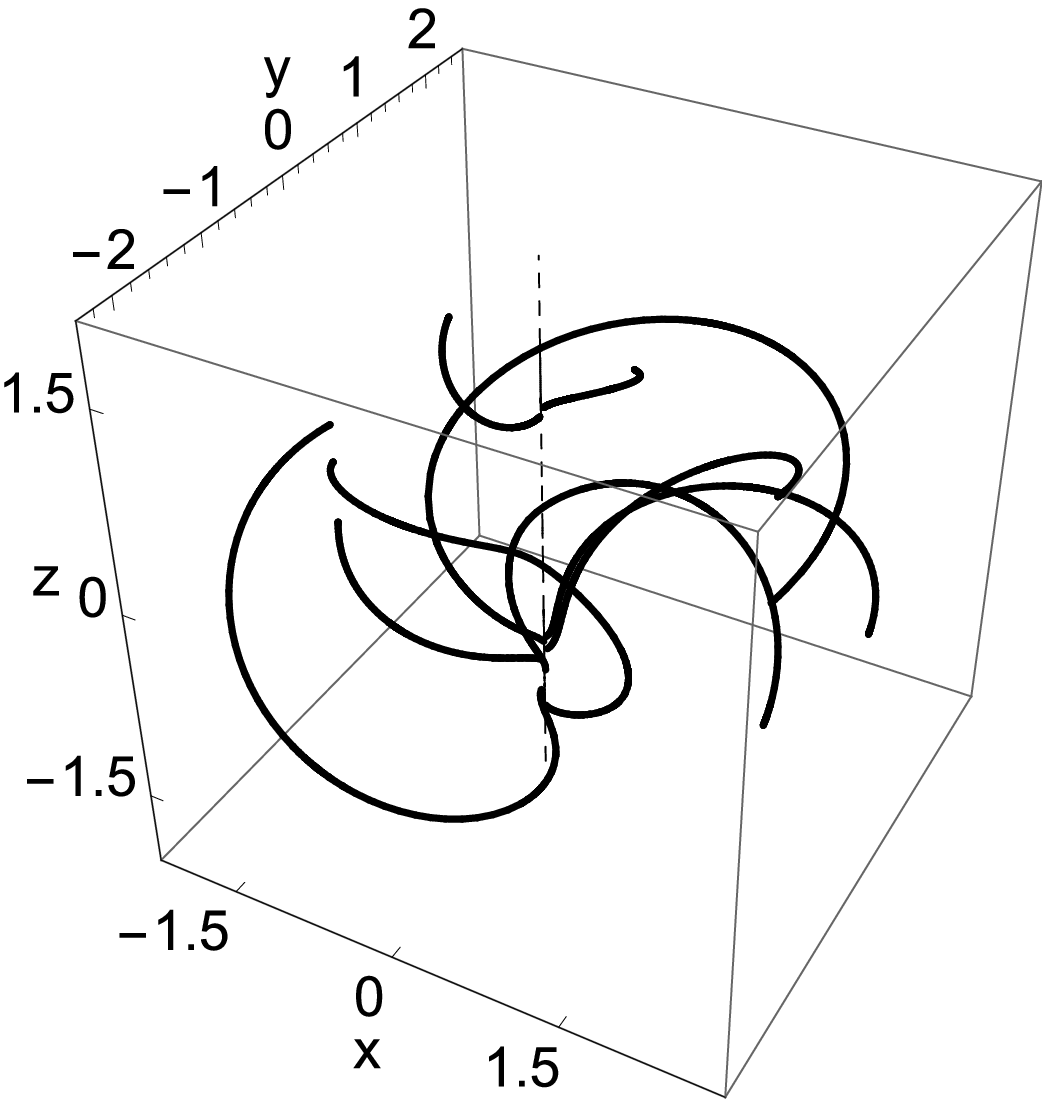}
\end{tabular}
\caption{(Color online) Orbits for massive test particles in presence of two KN black holes. (a) Circular orbit in $(\rho, z)$ coordinates, with parameters $R=3, M=a=Q=1, E=0.98$ and $L=10.48$. The dashed red curve represents the radii of the orbit ($\rho=46.06$). (b) Orbit (a) in $(x, y, z)$ coordinates. (c) Bounded orbit in $(\rho, z)$ coordinates, with parameters $R=3, M=a=Q=1, E=0.97$ and $L=8$. The dashed red curves represent the minimal and the maximal radii of the orbit ($\rho_{\rm min}=16.2$ and $\rho_{\rm max}=41.6$, respectively. (d) Orbit (c) in $(x, y, z)$ coordinates. (e) Oscillating orbit in  $(\rho, z)$ coordinates, with parameters $R=3, M=a=Q=1, E=0.9$ and $L=0$. The dashed red circle represents the amplitude of the orbit, $\rho=19$. (f) Orbit (e) in $(x, y, z)$ coordinates. (g) Plunging orbits in $(\rho, z)$ coordinates, with parameters $R=3, M=a=Q=1, E=0.97$ and $L=8$. (h) Orbit (g) in $(x, y, z)$ coordinates. In all the cases, the blue line represents the orbit in $(\rho, z)$ coordinates, and the curves in black the effective potential $V(\rho,z)$. The thick red lines at $\rho=0$ denote the KN black holes and the dashed black line the symmetry axis.} \label{fig7}
\end{figure}

\begin{center}
\begin{figure}
\begin{tabular}{cccc}
(a)&(b)&(c)&(d)\\
\includegraphics[width=4cm,angle=0]{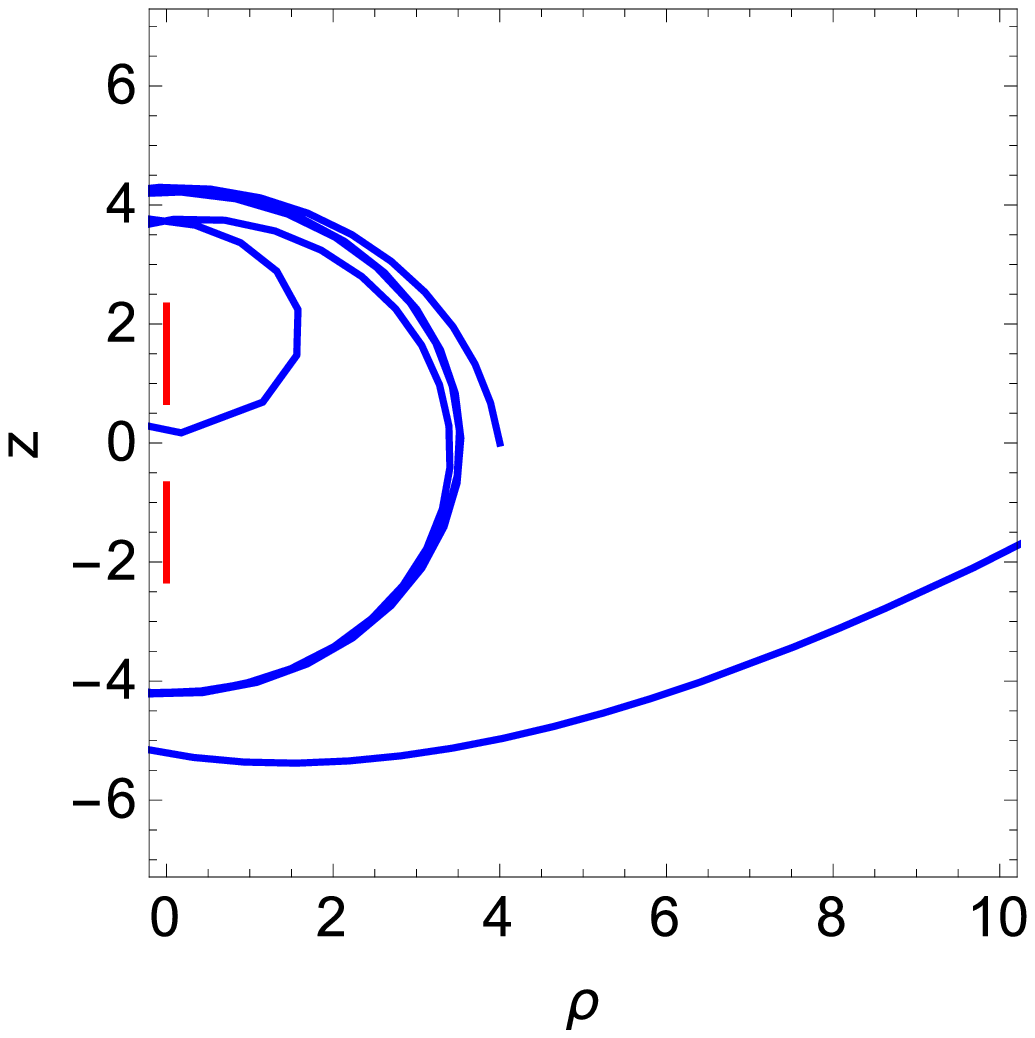}&
\includegraphics[width=4cm,angle=0]{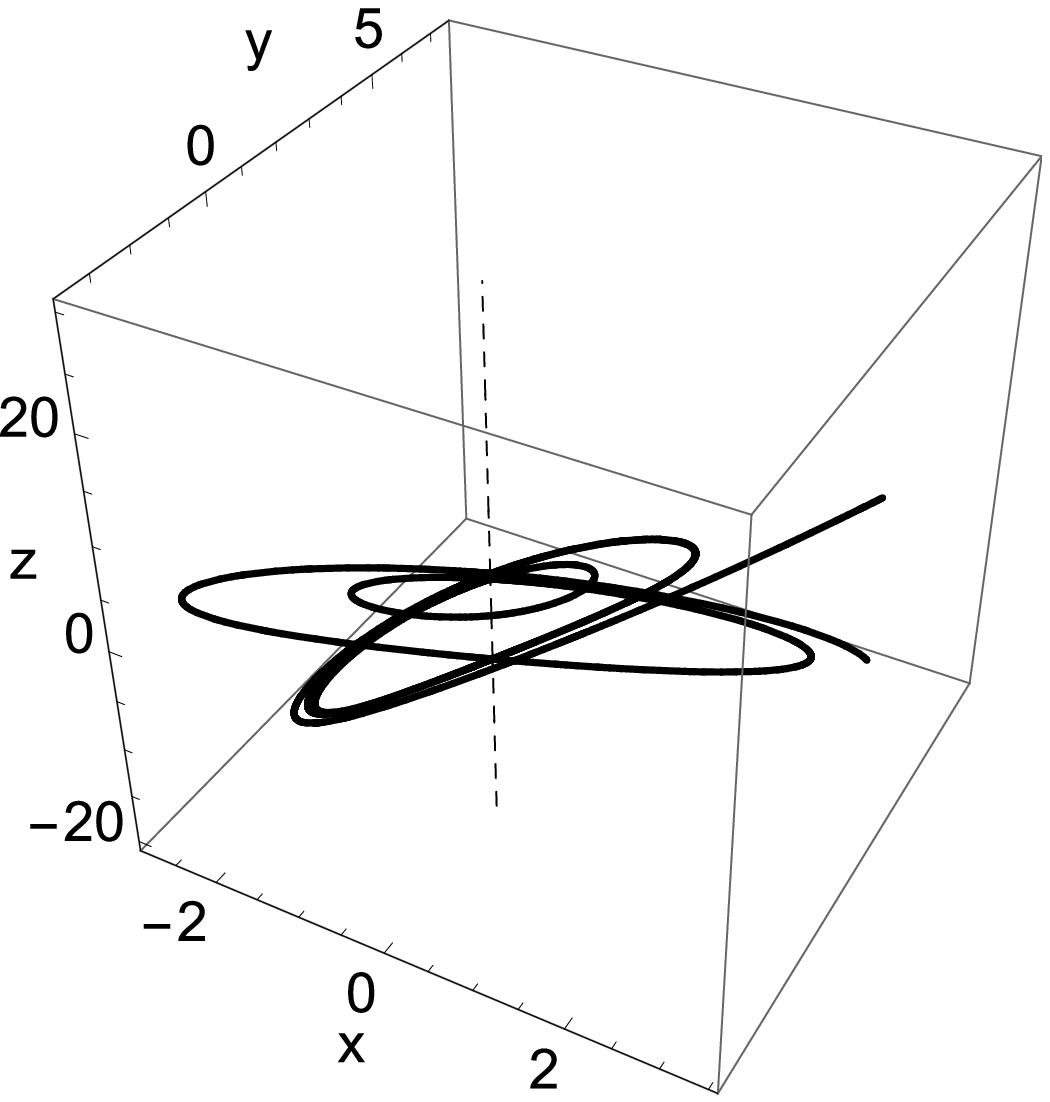}&
\includegraphics[width=4cm,angle=0]{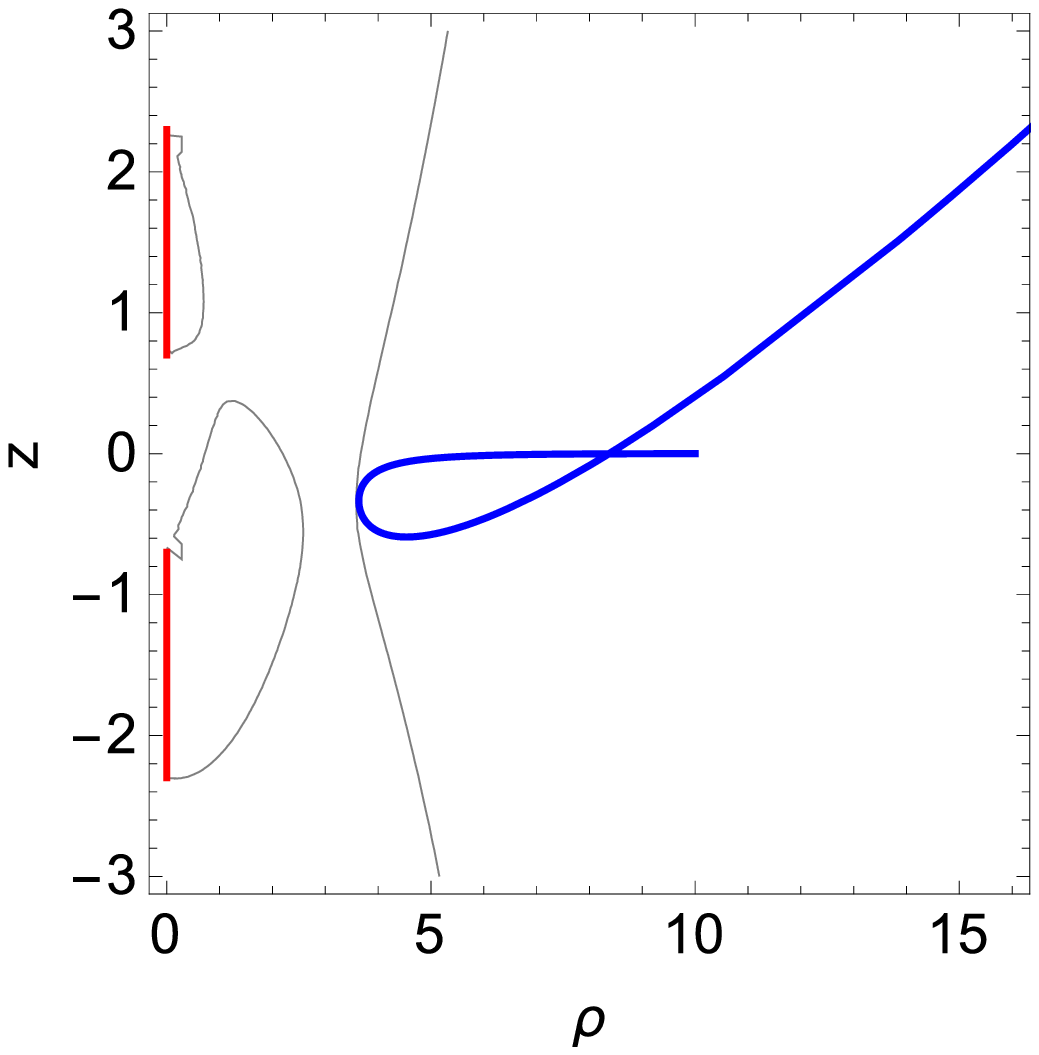}&
\includegraphics[width=4cm,angle=0]{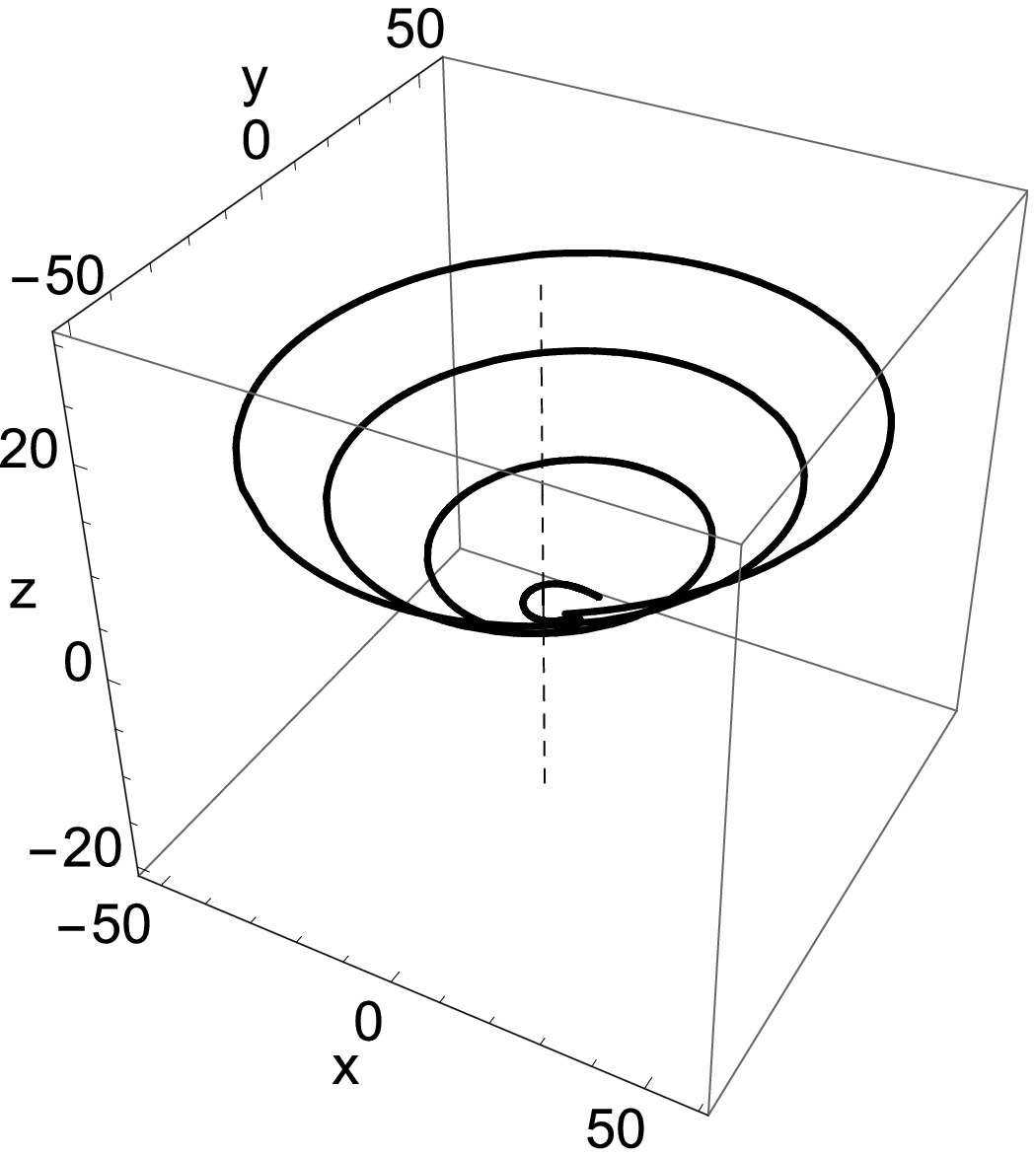}\\
(e)&(f)&(g)&(h)\\
\includegraphics[width=4cm,angle=0]{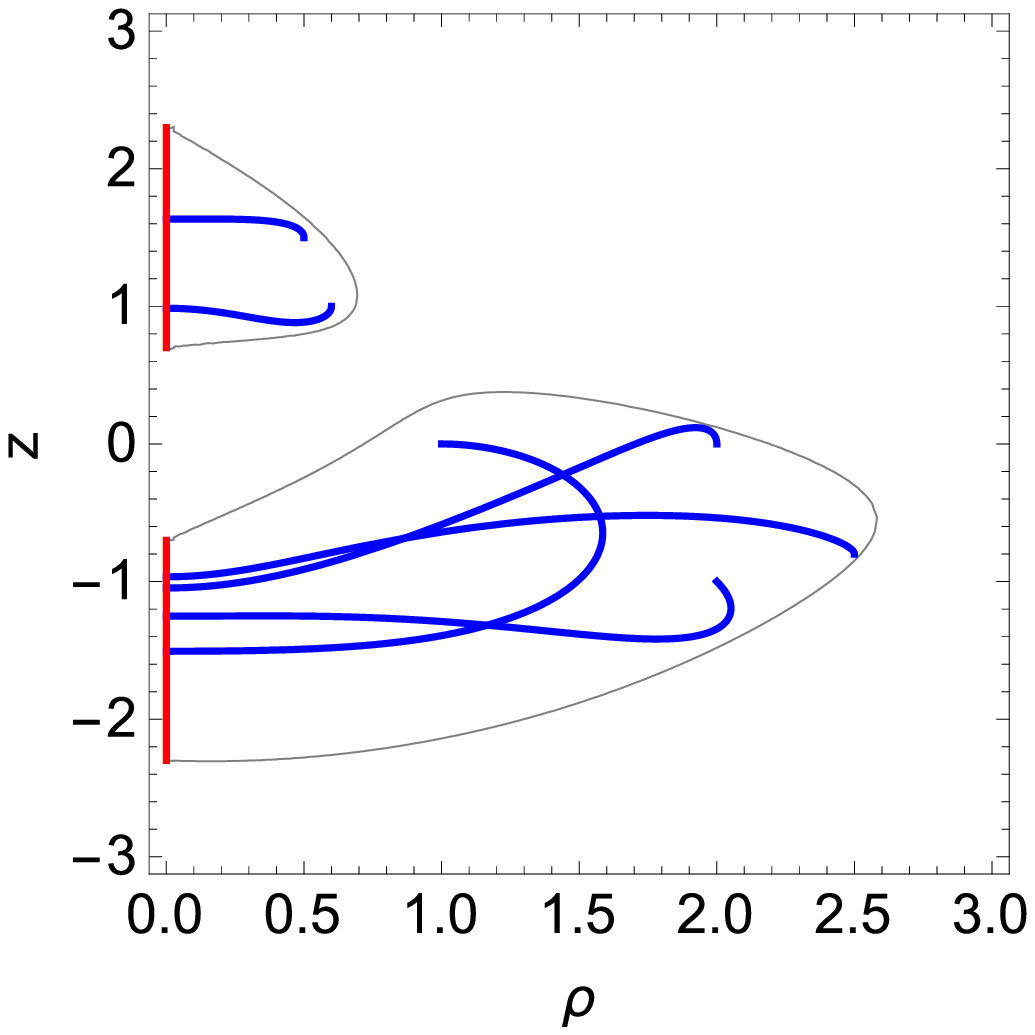}&
\includegraphics[width=4cm,angle=0]{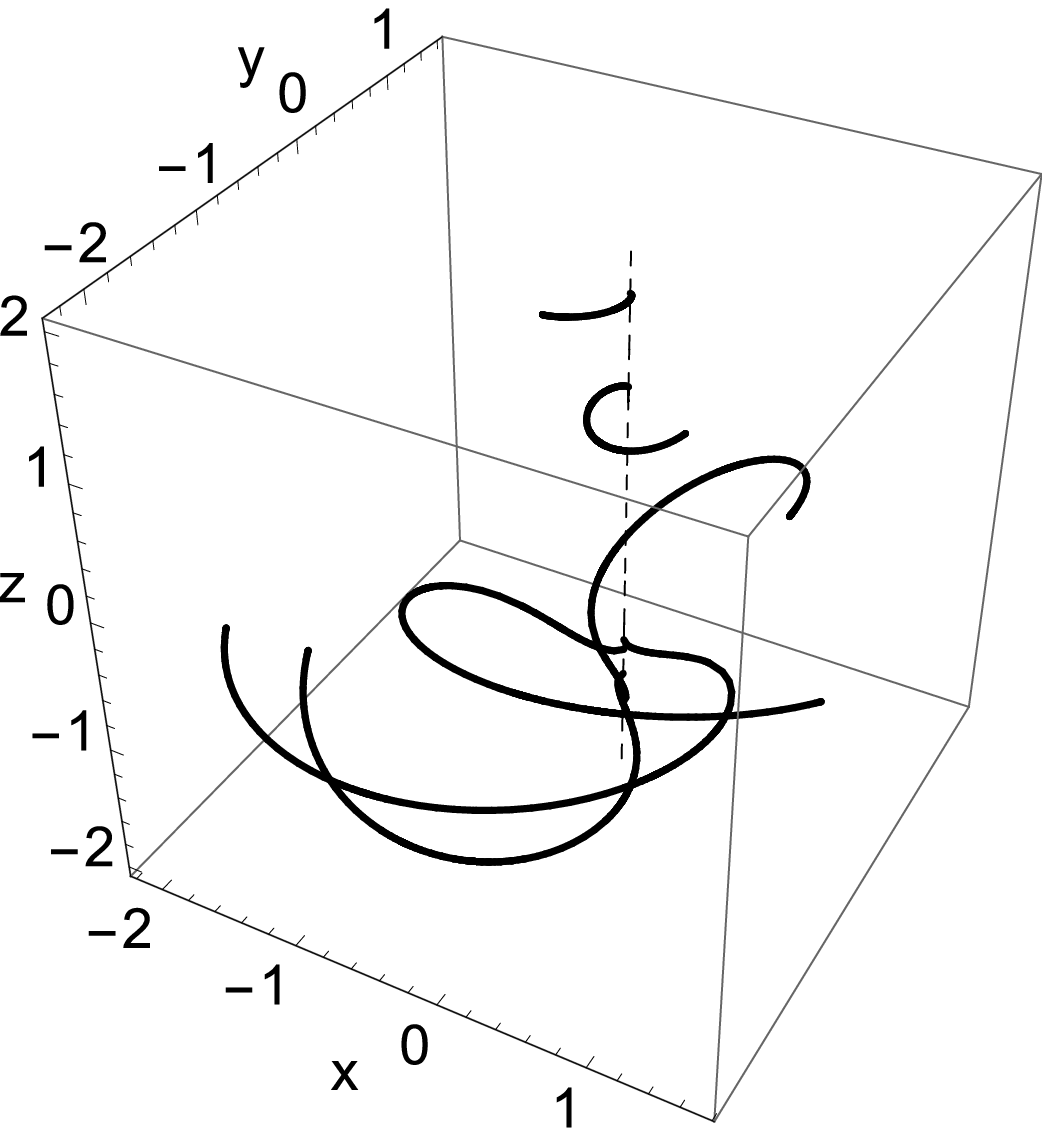}&
\includegraphics[width=4cm,angle=0]{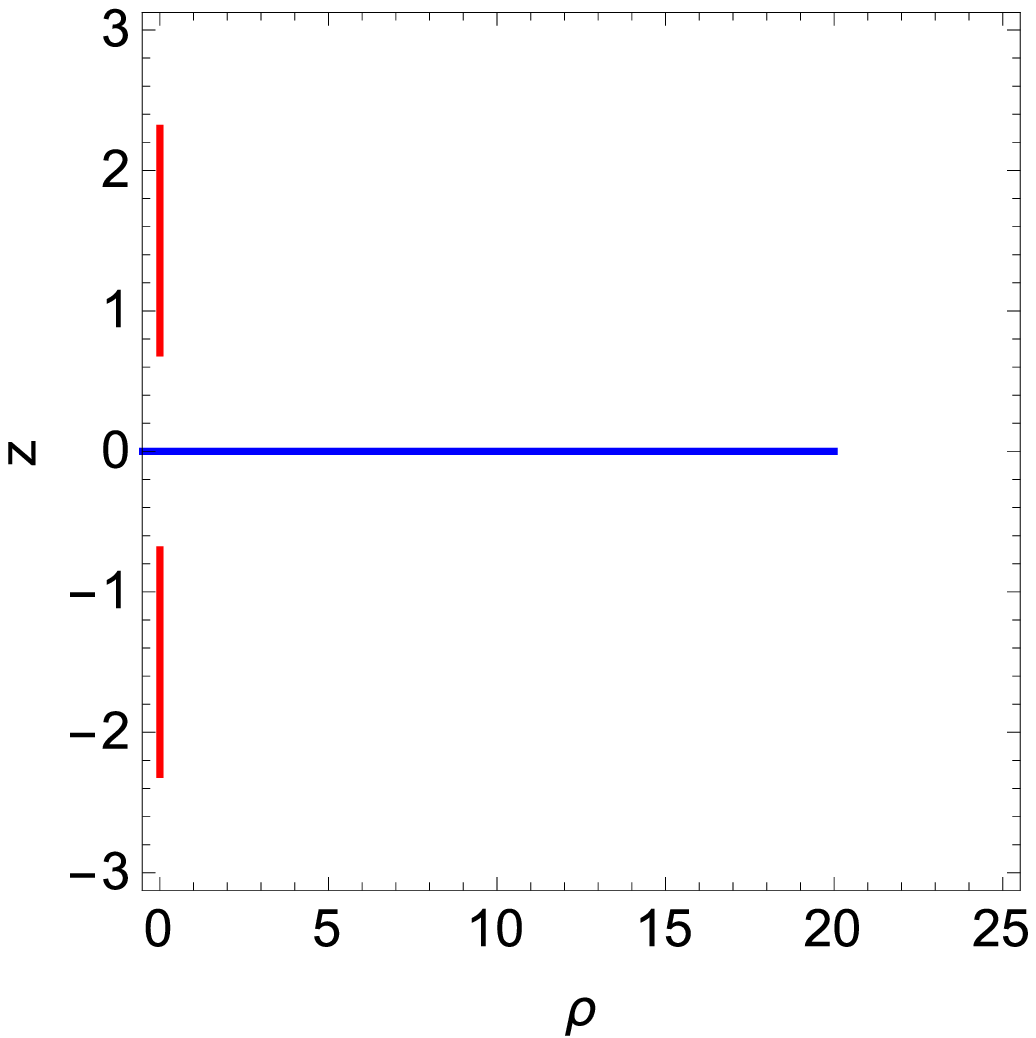}&
\includegraphics[width=4cm,angle=0]{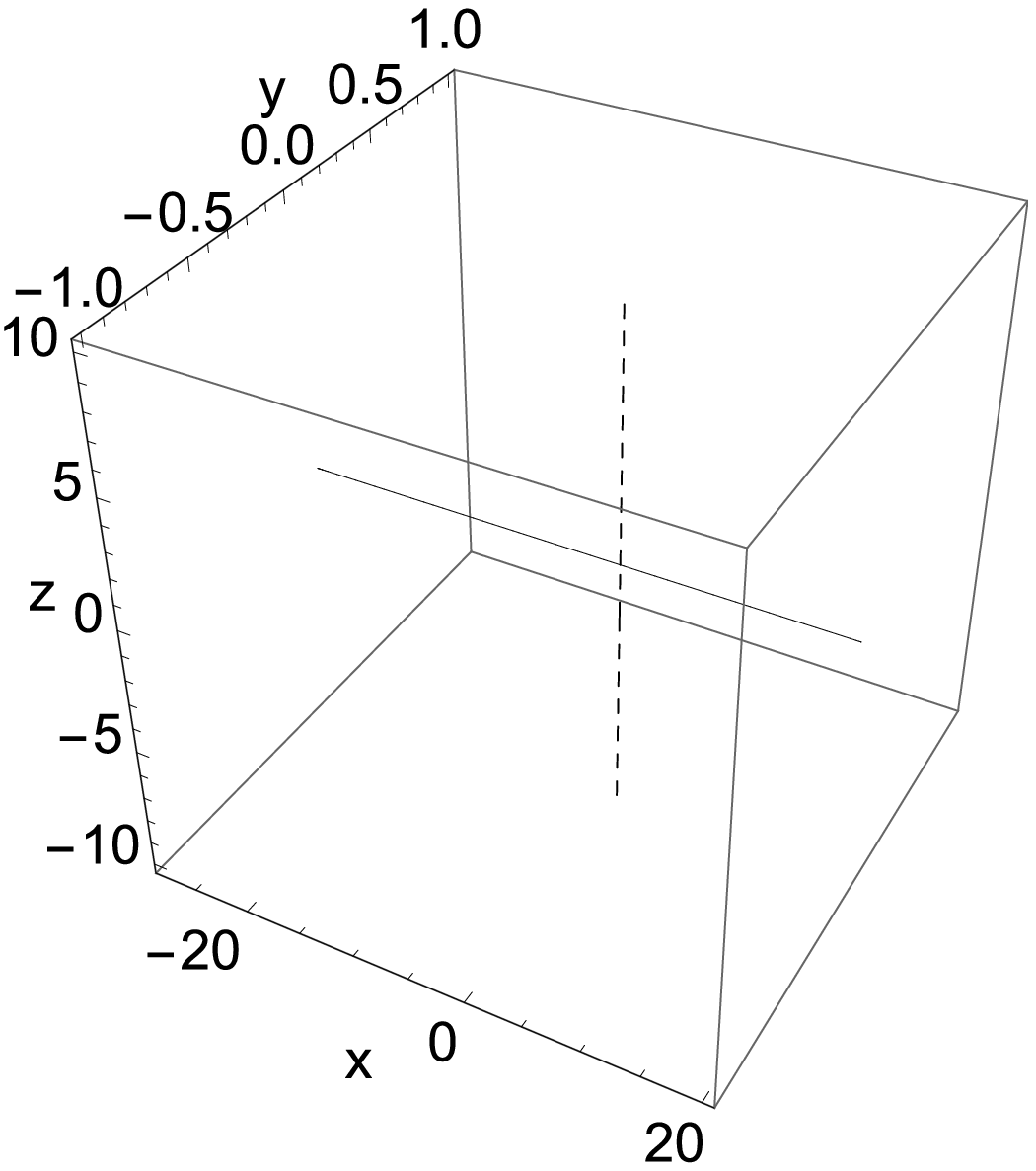}
\end{tabular}
\caption{(Color online) Orbits for photons in presence of two KN black holes. (a) Unstable orbit in $(\rho, z)$ coordinates, with parameters $R=3, M=a=Q=1, E=1$ and $L=0$. (b) Orbit (a) in $(x, y, z)$ coordinates. (c) Scattered orbit in $(\rho, z)$ coordinates, with parameters $R=3, M=a=Q=1, E=1.98$ and $L=20$. (d) Orbit (c) in $(x, y, z)$ coordinates. (e) Plunging orbits in  $(\rho, z)$ coordinates, with parameters $R=3, M=a=Q=1, E=1.98$ and $L=20$. (f) Orbit (e) in $(x, y, z)$ coordinates. (g) Unbounded orbit in $(\rho, z)$ coordinates, with parameters $R=3, M=a=Q=1, E=1$ and $L=0$. (h) Orbit (g) in $(x, y, z)$ coordinates. In all the cases, the blue line represents the orbit in $(\rho, z)$ coordinates, and the curves in black the effective potential $V(\rho,z)$. The thick red lines at $\rho=0$ denote the KN black holes and the dashed black line the symmetry axis.} \label{fig8}
\end{figure}
\end{center}
\end{widetext}

\end{document}